\documentclass{aa}  

\usepackage{thesis_defs}
\usepackage{graphicx}   
\usepackage[colorlinks=true,
linkcolor=blue,
citecolor=blue]{hyperref}
\usepackage{orcidlink}

\usepackage{txfonts}
\begin{document} 

\title{3D hybrid fluid-particle jet
simulations and the importance
of synchrotron radiative losses}

   \author{Joana A. Kramer
          \inst{1,2*} \orcidlink{0009-0003-3011-0454},
          Nicholas R. MacDonald\inst{3,1} \orcidlink{0000-0002-6684-8691},
          Georgios F. Paraschos\inst{1} \orcidlink{0000-0001-6757-3098}, Luca Ricci\inst{4,1} \orcidlink{0000-0002-4175-3194}}

   \institute{
   \inst{1}Max-Planck-Institut für Radioastronomie, Auf dem Hügel 69, D- 
                                53121 Bonn, Germany\\
       \inst{2}Theoretical Division, Los Alamos National Laboratory, Los Alamos, NM 87545, USA\\
                \inst{3}Department of Physics and Astronomy, 108 Lewis Hall,
University of Mississippi, Oxford, MS 38677-1848, USA\\
                \inst{4}Julius-Maximilians-Universität Würzburg, Institut für Theoretische Physik und Astrophysik, Lehrstuhl für Astronomie, Emil-Fischer-Str. 31, D-97074 Würzburg, Germany\\
                        \inst{*}\email{jah@lanl.gov}
             }

   \date{}

 \abstract
 {Relativistic jets in active galactic nuclei are known for their exceptional energy output, and imaging the synthetic synchrotron emission of numerical jet simulations is essential for a comparison with observed jet polarization emission. }
 {Through the use of 3D hybrid fluid-particle jet simulations (with the PLUTO code), we overcome some of the commonly made assumptions in relativistic magnetohydrodynamic (RMHD) simulations by using non-thermal particle attributes to account for the 
 resulting synchrotron radiation.  Polarized radiative transfer and ray-tracing (via the RADMC-3D code) highlight the differences in total intensity maps when (i) the jet is simulated purely with the RMHD approach, (ii) a jet tracer is considered in the RMHD approach, and (iii) 
 a hybrid fluid-particle approach
 is used. The resulting emission maps were compared to the example of the radio galaxy Centaurus A.}
 {We applied the Lagrangian particle module implemented in the latest version of the PLUTO code. This new module contains a state-of-the-art algorithm for modeling diffusive shock acceleration and 
 for accounting
 for radiative losses in RMHD jet simulations. The module implements the physical postulates missing in RMHD jet simulations by accounting for a cooled ambient medium and strengthening the central jet emission.}
 {We find a distinction between the innermost structure of the jet and the back-flowing material by mimicking the radio emission of the Seyfert II radio galaxy Centaurus\,A when considering an edge-brightened jet with an underlying purely toroidal magnetic field. We demonstrate the necessity of synchrotron cooling as well as the improvements gained when directly accounting for non-thermal synchrotron radiation via non-thermal particles. }
 {} 

   
  \authorrunning{J.~A.~Kramer et al.}
  \titlerunning{Ray-tracing in RMHD jet simulations}
 \maketitle
%
\section{Introduction}
Active galactic nuclei (AGN) are supermassive black holes (SMBHs) that actively accrete matter, forming highly energetic, bright, compact objects in the centers of galaxies~\citep{Rees}. Approximately 10\,\% of AGN are classified as radio-loud and are characterized by the presence of large-scale jets that remain collimated on up to kiloparsec scales~\citep{Homan}. Radio-loud AGN encompass various classes, including radio galaxies and blazars, which differ in orientation with respect to our line of sight. While radio galaxies and blazars are believed to be the same intrinsic object, blazars show jets that are increasingly aligned to our line of sight~\citep{Urry1995}.

The strength and orientation of magnetic fields in AGN can be determined through observations of power-law spectra and polarization signals in the radio regime. The strength of the intrinsic magnetic field is measured by assessing the apparent core shifts in optically thick synchrotron self-absorbed emission across various frequencies~\citep[see, e.g.,][]{Pushkarev2012}. \cite{Lobanov98} calculated the magnetic field strengths of various radio sources by analyzing the core shift. The direction of the magnetic field is inferred from polarization measurements, particularly from the detection of both linearly and circularly polarized non-thermal synchrotron emission~\citep{Meisenheimer, Carilli, Heavens, Brunetti}.

Astrophysical jets associated with AGN
are thought to carry helical magnetic field configurations near the central engine. However, the role of these helical fields on larger scales remains unclear \citep[e.g.,][]{Gabuzda2008,Gabuzda2018,Paraschos,Perucho2023,Kim2023}. The synchrotron emission from these jets exhibits high degrees of linear polarization~\citep[LP; up to 60$-$70\,\%; see, e.g.,][]{Rybicki79, Todorov, Lister2005, Park2020} and weaker circular polarization (CP; $\sim 1\,$\%). This polarized emission provides valuable insights into the intrinsic magnetic field structure~\citep[see, e.g.,][]{Tsunetoe}.  Numerical studies have shown that CP can be used to distinguish between poloidal and toroidal magnetic field morphologies within relativistic jets~\citep{Kramer&MacDonald}.

Multiwavelength observations, ranging from radio to high-energy gamma rays, offer valuable insights into the micro-physical processes occurring within AGN jets~\citep{M87multi}. These micro-physical processes~\citep{magrecon} occur on physical scales several orders of magnitude smaller than the overall jet scale. Bridging the gap between these scales is a significant challenge for theoretical models of AGN jet emission.

Due to the inability to reproduce relativistic jets in laboratories, numerical simulations are the primary means of exploring the underlying physics~\citep{Bellan}. AGN jets can be effectively modeled as plasma since the Larmor radius of the jet particles is much smaller than the spatial scales of the system~\citep{BlandfordRees}. Recently, a numerical algorithm has been developed within the PLUTO code to simulate multidimensional flow patterns, incorporating small-scale processes in a sub-grid manner~\citep{Mignone2007,Vaidya2018,Mignone2019, Mandal2023, Melon2019, Mignone2023}. 

In general, synchrotron emission signatures from large-scale jets are obtained through time-dependent simulations and post-process radiative transfer. In relativistic hydrodynamic contexts, transfer functions between thermal and non-thermal plasma are implemented. In the case of relativistic magnetohydrodynamic (RMHD) calculations, the internal magnetic structure of the jet and the parameters of the energy distribution are used to compute synchrotron emission maps~\citep{Porth2011, Mizuno2015, Fromm2016,Kramer&MacDonald}.

To study the micro-physics of particle acceleration at shocks, hybrid implementations that combine particle and grid-based fluid descriptions have been developed, targeting different scales of interest. The particle-in-cell approach is often employed to understand particle acceleration at relativistic shocks, particularly at the scales of the electron gyro-radius. Hybrid MHD-particle-in-cell approaches enable the study of shock acceleration at slightly larger length scales, typically of the order of a few thousand proton gyro-radii. These approaches describe the interaction between collisionless cosmic ray particles and thermal plasma, capturing small-scale kinetic effects in magnetosphere simulations~\citep[e.g.,][]{Bai, Marle, Mignone2018, Daldorff}. Recently, \cite{Dubey2023}, \cite{Mondal2023}, \cite{Nurisso2023}, and \cite{Girib} explored the behavior of the observable emission from particles encountering various shock acceleration mechanisms within the jet stream in hybrid (R)MHD-particle simulations. 

For relativistic hydrodynamic flows, the ``painting'' of  non-thermal particle (NTP) populations onto thermal flows of plasma allows the study of non-thermal synchrotron emission from internal shocks in blazars~\citep{Mimica2009,Mimica2012,Porth2011,Fromm2016}. In these models, a simplifying assumption was made by setting a constant value for the power-law index that governs particle injection at shocks ($N(E)\, \propto \, E^{-s}$), with values such as $s = 2.0$~\citep{delaCita}, $s = 2.23$~\citep{Fromm2016}, and $s = 2.3 $~\citep{Kramer&MacDonald}.

Fully Eulerian approaches to NTP transport have been successfully applied to numerical simulations over several decades~\citep[such as][]{Kang}.
Additional methods have been developed by~\cite{Vaidya2018} to overcome the aforementioned limitations and establish a state-of-the-art hybrid framework for particle transport. This framework aims to model high-energy non-thermal emission by including Lagrangian particles in large-scale 3D RMHD simulations. This is achieved by utilizing the magnetic field from the underlying RMHD simulation to calculate radiative losses, such as synchrotron cooling. The hybrid framework can be compared with observations by incorporating micro-physical aspects of spectral evolution based on local fluid conditions.

The emission from the accelerated charged particles within jets in various types of AGN, for example  blazars and Seyfert II radio galaxies such as Centaurus A, is relativistically Doppler-beamed. Relativistic plasma and magnetic fields in AGN jets give rise to incoherent synchrotron emission components ranging in wavelength from the radio to the optical, the UV, or even X-rays, and can exhibit LP. The polarization parameters carry information about the jet's physical conditions, including the magnetic field strength, topology, and particle density~\citep{Wardle}.

This paper explores the effect that synchrotron cooling has on ray-traced images of a 3D RMHD jet with an orientation similar to Centaurus A. In particular, synchrotron cooling helps highlight the jetted structure relative to emission from the back-flowing cocoon of the jet (which has undergone synchrotron cooling). Previous models utilizing the PLUTO code have relied on numerical tracer values within the jet to artificially exclude the bow shock from emission calculations~\citep[as demonstrated in][]{Kramer&MacDonald}. This paper demonstrates that incorporating particle physics is crucial for accurately modeling extragalactic sources (in the radio domain). We aim to mimic the characteristics of the nearby radio galaxy Centaurus A, and we discuss a qualitative comparison between our synthetic polarized emission maps and observations obtained from both the TANAMI program~\citep{muller2014} and the Event Horizon Telescope Collaboration (EHTC), under the leadership of \cite{JanssenCenA}.
A similar approach to mimic the radio galaxy 3C\,84, using the methods described in this paper, is also successfully implemented in an accompanying paper \citep{Paraschos24}.

This paper is organized as follows: In Sect.~\ref{sec:2} we provide an introduction to the principles of the hybrid fluid-particle simulations used in the PLUTO code, with a focus on Lagrangian particles (i.e., macro-particles that represent numerous non-thermal micro-particles). We introduce the principles of particle physics within numerical jet simulations, namely the inclusion of radiative losses. We further focus on the numerical implementation of our relativistic jet simulations, along with the explanation of the post-process generation of a 3D grid, which is extrapolated from the NTP attributes. 
Section~\ref{sec:3} outlines our applied method for calculating synthetic polarized synchrotron emission maps from particle simulations of jets via the ray-tracing code RADMC-3D. Section~\ref{sec:4} compares our simulations to the nearby radio galaxy Centaurus A.
Section~\ref{sec:5} discusses the comparison between our numerical hybrid fluid-particle RMHD approach and observations of Centaurus A~\citep{muller2014,JanssenCenA}. Finally, Sect.~\ref{sec:6} provides our conclusions.

\section{Lagrangian module}\label{sec:2}
In recent years, researchers have recognized the importance of incorporating Lagrangian particles into (R)MHD simulations to gain deeper insights into the behavior of the jet plasma~\citep{Vaidya2018}. Lagrangian particles are individual tracer elements that move with the thermal jet flow, providing valuable information about the fluid's characteristics, such as its velocity, density, and magnetic field strength. By following the trajectories of these particles, we can track the transport and mixing of different plasma components. In particular, we examined the impact of magnetic fields on particle dynamics and studied various physical processes occurring within the jet.
The use of Lagrangian particles in RMHD jet simulations allows us to perform a more detailed analysis of the complex phenomena taking place in these astrophysical systems. Hence, this approach enables the study of important processes, including particle acceleration mechanisms, the formation and propagation of shocks, and the effect of radiative losses within the relativistic jet.

\subsection{Cosmic ray transport equation} 
The reduced cosmic ray transport equation for the relativistic case~\citep{Webb} considers particle transport by both convection and diffusion, along with their evolution of momentum space. This includes energy changes due to adiabatic expansion or contraction, as well as caused by radiation processes. 
In particular, the time evolution of a NTP is 
governed by the 
following expression~\citep{Vaidya2018}:
\begin{align}\label{5.1}
\frac{\text{d}n_\mathrm{e}(E)}{\text{d}\tau} + \frac{\partial}{\partial 
        E}\left[\left(-\frac{E}{3}\nabla_\mu u^\mu + 
\dot{E}_l\right)n_\mathrm{e}(E)\right] = -n_\mathrm{e}(E) 
\nabla_\mu u^\mu,
\end{align}
where $u^\mu$ is the bulk four-velocity, d$\tau$ is a time differential (related to the laboratory time), and $\gamma$ is the electron Lorentz factor of the NTP. The $n_\mathrm{e}\left(E\right)$ represents an electron (subscript e) energy distribution normalized to the particle number density. The particle energy distribution is initialized as a single power-law, with index $s$, in the form of $n_\mathrm{e}\left(E\right) \propto\,  N_\mathrm{e} E^{-s}$. The terms within Eq.~\ref{5.1} in square brackets account for energy changes due to adiabatic expansion and radiative losses.
This paper focuses on the second term ($\dot{E}_l$), which accounts explicitly for synchrotron cooling~\citep{Vaidya2018}.

Equation~\ref{5.1} is solved in two steps. In the first step, the particle positions 
$\VEC{x}_p$ are evolved in time according to
\begin{align}\label{eq:xp}
\frac{\text{d}\VEC{x}_p}{\text{d}t} = \VEC{v}_p = \VEC{v}(\VEC{x}\rightarrow 
\VEC{x}_p),
\end{align}
where $\VEC{v}_p$ is the particle velocity. The solution is performed 
        using a 
        second-order Runge-Kutta scheme that computes a  predictor step followed by a corrector step~\citep[see, e.g., ][]{Mignone2007,Vaidya2016,Vaidya2018, Mignone2019}.
In the second step, the energy change is accounted for by updating the energy distribution of the particle. Consequently, the energy term within Eq.~\ref{5.1} is solved. Section~\ref{sec:synchr} provides a more detailed description of this solution considering radiative synchrotron losses in particular~\citep{Vaidya2018}.

\subsection{Synchrotron losses}\label{sec:synchr}
To trace the synchrotron losses, the energy distribution of each macro-particle is continuously updated over time. Precisely, the code keeps track of the particle's parameters at a given position on the Cartesian grid (by tracking the particle position according to Eq.~\ref{eq:xp}).
The particle attributes are transferred in each numerical step and solved for the particle's energy changes. The attributes are computed in the following fashion:
\begin{align}\label{eq:energyloss}
    \frac{\mathrm{d}E}{\rm{d}\tau_\mathrm{p}} = -c_1\left(\tau_\mathrm{p}\right)E -c_2\left(\tau_\mathrm{p}\right)E^2 \equiv \Dot{E}.
\end{align}
In~Eq.~\ref{eq:energyloss}, the first term represents the energy loss resulting from adiabatic expansion, while the second term represents the combined energy loss from synchrotron radiation and inverse Comp-
ton scattering off the cosmic microwave background\footnote{It is assumed that scattering in the rest frame of relativistic particles follows the Thompson scattering model and the cosmic background radiation is isotropic}. The physical constants $c_1$ and $c_2$ are given as follows~\citep{Vaidya2018}:
\begin{align}
    \begin{split}
        c_1 = \frac{\nabla_\mu u^\mu}{3}, \qquad
        c_2 = \frac{4\sigma_\mathrm{T}c\beta^2}{3 m_e^2 c^4}\left[U_\mathrm{B} + U_\mathrm{rad}\left(E_\mathrm{ph}\right) \right].
    \end{split}
\end{align}
The Thomson cross-section, denoted as $\sigma_\mathrm{T}$, represents the scattering efficiency. The quantities $U_\mathrm{B}$ and $U_\mathrm{rad}$ correspond to the energy densities of the magnetic field and the radiation field, respectively. $E_\mathrm{ph}$ represents the energy of the incident photon field from the cosmic microwave background~\citep{Vaidya2018}.

\subsection{Numerical implementation}
The Lagrangian macro-particles represent 
ensembles of NTPs with finite energy 
distributions that are advected along fluid streamlines. The underlying flow is calculated by the RMHD module within PLUTO. Details on the principles of the RMHD module and physical scaling are provided in~\cite{Kramer&MacDonald}, based on~\cite{Mignone2007}. The implementation of the magnetic field remains the same as in~\cite{Kramer&MacDonald}. This paper focuses on a purely toroidal magnetic field prescription within a defined magnetization radius within to relativistic jet (i.e., $\sigma_\phi=1$). 

PLUTO stores the normalized energy distribution $n_\mathrm{e}\left(E\right)$ of NTPs in logarithmically separated energy bins $E_\mathrm{min}\leq E \leq E_\mathrm{max}$ (with $E=E_\mathrm{bin}$ in units of electron rest mass). We assumed an 
initial energy distribution at time step 0 ($n_\mathrm{e}^0\left(E\right)$), of the NTPs that follows a global power-law distribution in energy, similar to~\cite{Vaidya2018}:
\begin{align}\label{eq:powerlaw}
 n_{\mathrm{e}}^0\left(E\right) = n_{\mathrm{micro}}\left(\frac{1-s}{E_{\mathrm{max}}^{1-s}-E_{\mathrm{min}}^{1-s}}\right)E^{-s}.
\end{align}
We initialized the spectrum with a power-law index of $s=2$. The spectral index, $\alpha$, which is needed to calculate the resulting synchrotron emission,  relates to the power-law index $s$ via: $\alpha = (s-1)/2$. The power-law index and the spectral index are computed for each cell and, hence, are changing constantly throughout the computational domain (both spatially and temporally). \footnote{We computed spectral index ``cubes'' that are input into RADMC-3D for the post-process ray-tracing.} n$_\mathrm{micro}$ denotes the initial number density of physical particles across energy bins: 
\begin{align}\label{eq:6}
  n_\mathrm{micro} \equiv \int_{E_\mathrm{min}}^{E_\mathrm{max}} n_\mathrm{e}\left(E\right) \mathrm{d}E.
\end{align}
This formulation calculates the total energy in dimensionless units, that is, the area under the distribution per macro-particle in the energy spectra illustrated in Fig.~\ref{fig:spectrum}. To account for physical values, $n_\mathrm{micro}$ is expressed in terms of density per atomic mass unit. We would like to emphasize that the expression in Eq.~\ref{eq:6} changes for every Lagrangian particle, which is affected by the synchrotron cooling term in Eq.~\ref{5.1}.  

To calculate the synthetic non-thermal synchrotron emission from the Lagrangian particle attributes, we rewrote the power-law in terms of the electron Lorentz factor,\footnote{The Lorentz factor equals the energy bins in the numerical framework.} $\gamma$ ($E=\gamma m_e c^2$). Similar to the study conducted in  \cite{Kramer&MacDonald}, we set an initial bound between the lower energy cutoff $\gamma_\text{min}=10$ and the upper limit $\gamma_\text{max}=\gamma_\text{min}\cdot10^6$. Similar to \cite{Kramer&MacDonald}, we generated 3D grids of magnetic field strength $\textbf{B}$ [Gauss], electron number densities $n_\mathrm{micro}$ [cm$^{-3}$], low-energy power-law cutoffs $\gamma_\mathrm{min}$, and spectral indices $\alpha$, from interpolation of our Lagrangian particle properties. These interpolated grids are then input into RADMC-3D. In order to compute the normalization constant, we further expressed the electron power-law distribution in terms of the electron Lorentz factor, $\gamma$: $n_e(\gamma) = N_\mathrm{e} \left(\frac{\gamma}{\gamma_\mathrm{min}}\right)^{-s}$ for $\gamma_\mathrm{min} \leq \gamma \leq \gamma_\mathrm{max}$. Together with the continuous equivalent of Eq.~\ref{eq:powerlaw}, we arrive at the normalization constant
\begin{align}\label{eq:norm}
    N_\mathrm{e} = \frac{\left(1-s\right) n_\mathrm{micro}}{\gamma_\mathrm{min}^s \left(\gamma_\mathrm{max}^{1-s}-\gamma_\mathrm{min}^{1-s}\right)}.
\end{align}
The particle attributes, such as $n_\mathrm{e}\left(\gamma\right)$, were evolved during the simulation.
In a post-process step, we estimated the normalization constant, $N_\mathrm{e}$, from the lower energy cutoff$,\gamma_\mathrm{min} \left(\equiv E_\mathrm{min}\right)$, and the power-law index, $s$, between two neighboring energy bins at a specific electron energy, $\gamma,$ that corresponds to a critical synchrotron frequency that we specify. This energy corresponds to our desired frequency, $\nu_\mathrm{obs}$, in the radio regime. The values for $N_\mathrm{e}$, $\gamma_\mathrm{min}$, and $s$ (i.e., $\alpha$) for each particle at a specific epoch within the simulation were estimated and interpolated onto a 3D Cartesian grid (details are described in Sect.~\ref{sec:2.4}).

At this point, we applied a physical scaling to the dimensionless grid values:
\begin{align}
\begin{split}
\rho_\text{cgs} &= \rho \cdot \rho_0,\\
p_\text{cgs} &= p \cdot \rho_0v_0^2,\\
\VEC{B}_\text{cgs} &= \VEC{B}\cdot \sqrt{4\pi\rho_0v_0^2},
\end{split}
\end{align}
where $\rho$, $p$, and $\VEC{B}$ denote the dimensionless PLUTO values. For all our 
simulations we specified the following unit values: $\rho_0 
\simeq 
1.0\cdot\SI{e-21}{\gram\,\centi\meter^{-3}}$, $L_0 \simeq 
3.0857\cdot\SI{e16}{\centi\meter}$, and $v_0\simeq 
2.998\cdot\SI{e10}{\centi\meter\per\second}$. With this choice of values, the magnetic field strength along the jet will be on Gauss to milli-Gauss scales. $n_\mathrm{micro}$ scales as $[\rho_0/ \,m_\mathrm{u}]$, with the atomic mass unit $m_{\mathrm{u,cgs}} \simeq 1.661\cdot 10^{-24}\,$g.
The jet radius, 
$r_j=\sqrt{\left(x-x_c\right)^2+\left(y-y_c\right)^2}$, denotes the radius of the particle injection zone. For the initialization of the injection of the particles, we applied an initial power-law index of $s=2$ (i.e., a spectral index of $\alpha=0.65$) and $n_\mathrm{micro}=0.001$ (in nondimensional code units) before these parameters evolved with the simulation. The choice of power-law index was motivated by observations of AGN jets~\citep[see, e.g.,][]{Paraschos}.

\subsection{Post-processing}\label{sec:2.4}

The generation of (polarized) synthetic synchrotron emission maps using the software RADMC-3D requires the normalization constants, $N_\mathrm{e}$, the lower-energy cutoffs, $\gamma_\mathrm{min}$, and the spectral indices, $\alpha$ (i.e., related to the power-law index, $s$) from the NTPs energy distribution, $n_\mathrm{e}\left(\gamma\right)$. However, radiative losses such as synchrotron cooling influence the high-energy tail of the initial power-law distribution.

\begin{figure}
    \centering
\includegraphics[trim={0 0 1cm 0}, clip,width=0.49\textwidth]{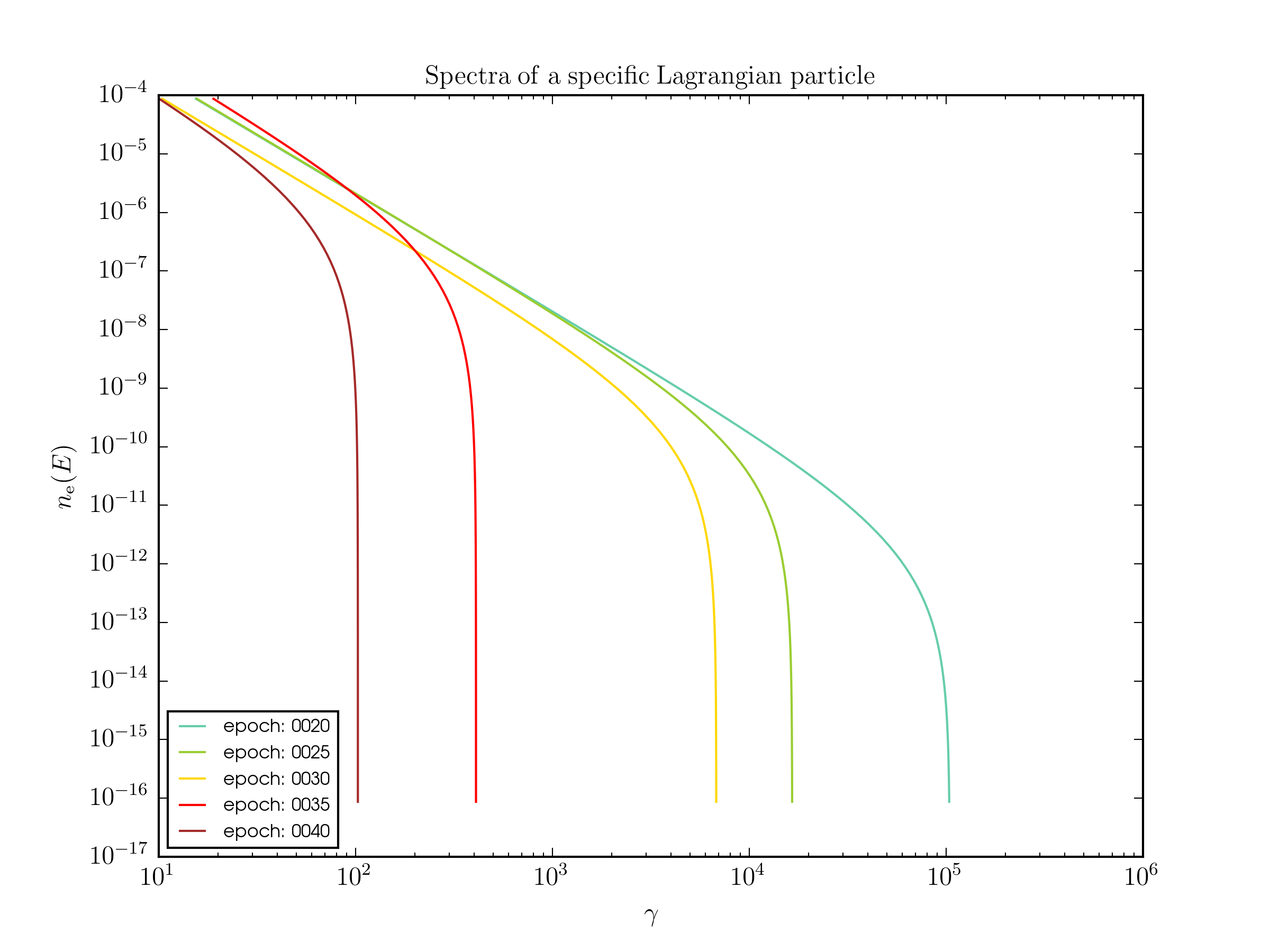}
    \caption{Normalized spectral distribution in dimensionless grid units of a representative macro-particle. The corresponding Lagrangian particle is located in the jet's spine just before entering the back-flow. The initial energy bins are chosen to be within $10\leq \gamma \leq 10^7$. The particle is injected onto the grid through the jet orifice at epoch 20 (each color represents a different simulation epoch). The radiative losses experienced by this NTP are visible as the particle's energy distribution evolves (in time) from right to left.}
    \label{fig:spectrum}
\end{figure}
Figure~\ref{fig:spectrum} shows the spectrum (in energy) of a Lagrangian macro-particle for different simulation epochs. The spectrum mainly mimics the energy changes due to adiabatic losses and synchrotron cooling\footnote{We would like to emphasize that we numerically excluded shock acceleration and that, hence, the spectrum displays energetic losses only.}. The particle starts to appear on the grid at epoch $20$ (mint-colored, on the far right) and loses energy while moving along the jet stream and eventually into the back flow. The energy loss is more pronounced when leaving the jet flow in epoch $35$ (red, second from left), after which the particle continues to ``cool'' into epoch $40$ (brown, most left). The particle in Fig.~\ref{fig:spectrum} represents a Lagrangian particle that has just left the jet stream, located in the jet lobe.

Across time steps, the particle undergoes the effect of synchrotron cooling, also referred to as synchrotron aging. The left part of the energy distribution can be described by the power-law distribution in Eq.~\ref{eq:powerlaw}. However, the right tail steepens due to radiative losses (Fig.~\ref{fig:spectrum}). We focused our attention on the energetic particles within the spine of the jet.
To compute synchrotron emission from our jet simulations, we computed the synchrotron spectral index (on a cell-by-cell basis) based on interpolations of local NTP power-law indices, the slope of which can be calculated between energy bins. This was done in adjacent energy ranges that result in the desired frequency of emission (accounting for Doppler beaming). For the sake of clarity, we plot the geometric mean of the energy of each bin on the x-axis and the corresponding value on the y-axis. This approach makes sense because the 100 bins would otherwise be very narrow. Additionally, we tested and confirmed that increasing the number of bins does not alter the results.

Our procedure ignores macro-particles that cooled ``away'' due to synchrotron aging at the specific observational frequency, namely the energy that we chose for our post-process calculations.

To determine the energy bin in terms of $\gamma$ that corresponds to an observational frequency, $\nu_\mathrm{obs}$, in our synthetic polarized maps in the radio band, we computed the critical frequency (in the emitted frame of the plasma, $\nu_\mathrm{em}$) at which most of the synchrotron power is emitted, as a function of each plasma cell's magnetic field strength, $B$, which is given by~\citep{nrao}\begin{align}\label{eq:prop}
    \nu_\mathrm{em} \,\sim\, \frac{\gamma^2 e B}{2 \pi m_e c},
\end{align}
where $e$ is the electron charge. 
We accounted for Doppler beaming when moving from the emitter's frame:
\begin{align}\label{eq:doppler}
    \nu_\text{obs} = \frac{\nu_\text{em}}{\Gamma\left(1-\beta \cos{\theta_\text{obs}}\right)},
\end{align}
where $\nu_\text{obs}$ is the observed frequency, and $\nu_\text{em}$ the emitted frequency. This is a function of the relativistic velocity $\beta=\nicefrac{v}{c}$, and the observed viewing angle $\theta_\mathrm{obs}$. We chose a bulk Lorentz factor of $\Gamma=7.088$.

\begin{figure}
    \centering
    \includegraphics[trim={14cm 4cm 19cm 0}, clip,width=0.49\textwidth]{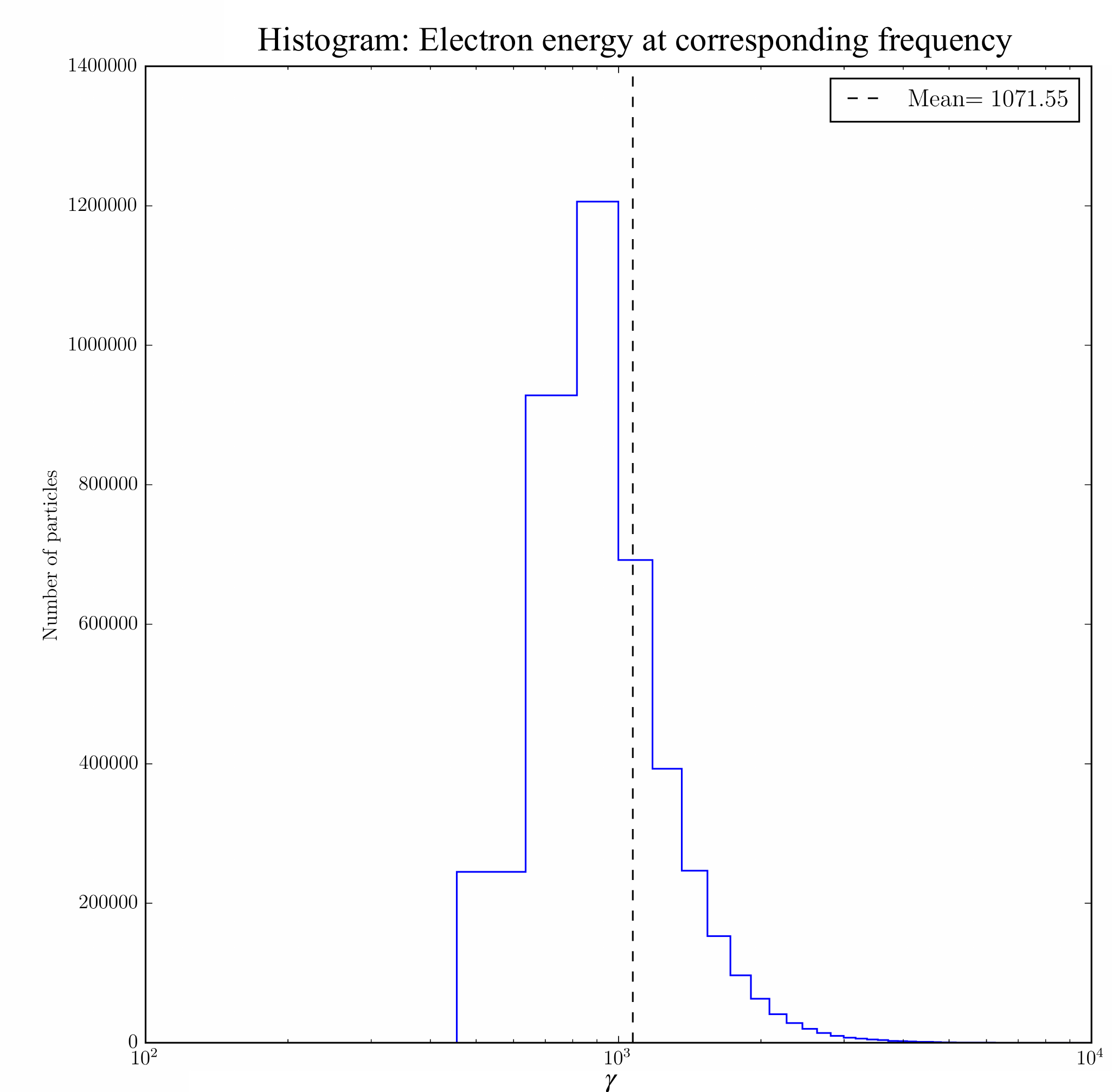}
    \caption{Number of particles per energy in dimensionless units\protect\footnotemark  corresponding to an observational frequency of $\nu_\mathrm{obs}=86$\,GHz. The mean energy bin, denoted by the dashed vertical line, excludes a large number of particles (that have cooled out of the range of interest) from our emission calculations.}
    \label{fig:hist}
\end{figure}

Equation~\ref{eq:prop} depends on the intrinsic magnetic field within the PLUTO simulation at each macro-particle's positions.
Hence, $E$ at $\nu_\mathrm{obs}$ differs slightly for each macro-particle, as shown in Fig.~\ref{fig:hist}. 
The histogram represents the number of particles that have energy distributions resulting in synchrotron emission peaked at a frequency of $\nu_\mathrm{obs}=86\,$GHz. For simplicity,
we used the mean value of these bins, namely, $\gamma =1071.55$, for the estimation of the spectral index $\alpha$. This step removes  from our emission calculations the 48\% of NTPs that have cooled beyond our desired synchrotron frequencies.
 
\footnotetext{The energy bin represents the electron Lorentz factor when applying electron rest mass energy units.}

Finally, our post-processing analysis results in values for $N_\mathrm{e}$, $\gamma_\mathrm{min}$, and $\alpha$ at irregular macro-particle positions within the jet flow of our hybrid fluid-particle simulation. We then interpolated these values to create a continuous 3D Cartesian grid. This grid is used by RADMC-3D to generate synthetic synchrotron emission maps. Our post-processing enhances our synthetic polarized maps, making them more comparable to observational data. The variable spectral index accounts for cooling effects visible when comparing different observational frequencies (illustrated in Appendix~\ref{app:freq}). In particular, we computed a variable spectral index cube that represents different spectral indices in different regions within the jet.

\begin{figure}
    \centering
    \includegraphics[trim={21cm 1cm 13cm 0}, clip,width=0.49\textwidth]{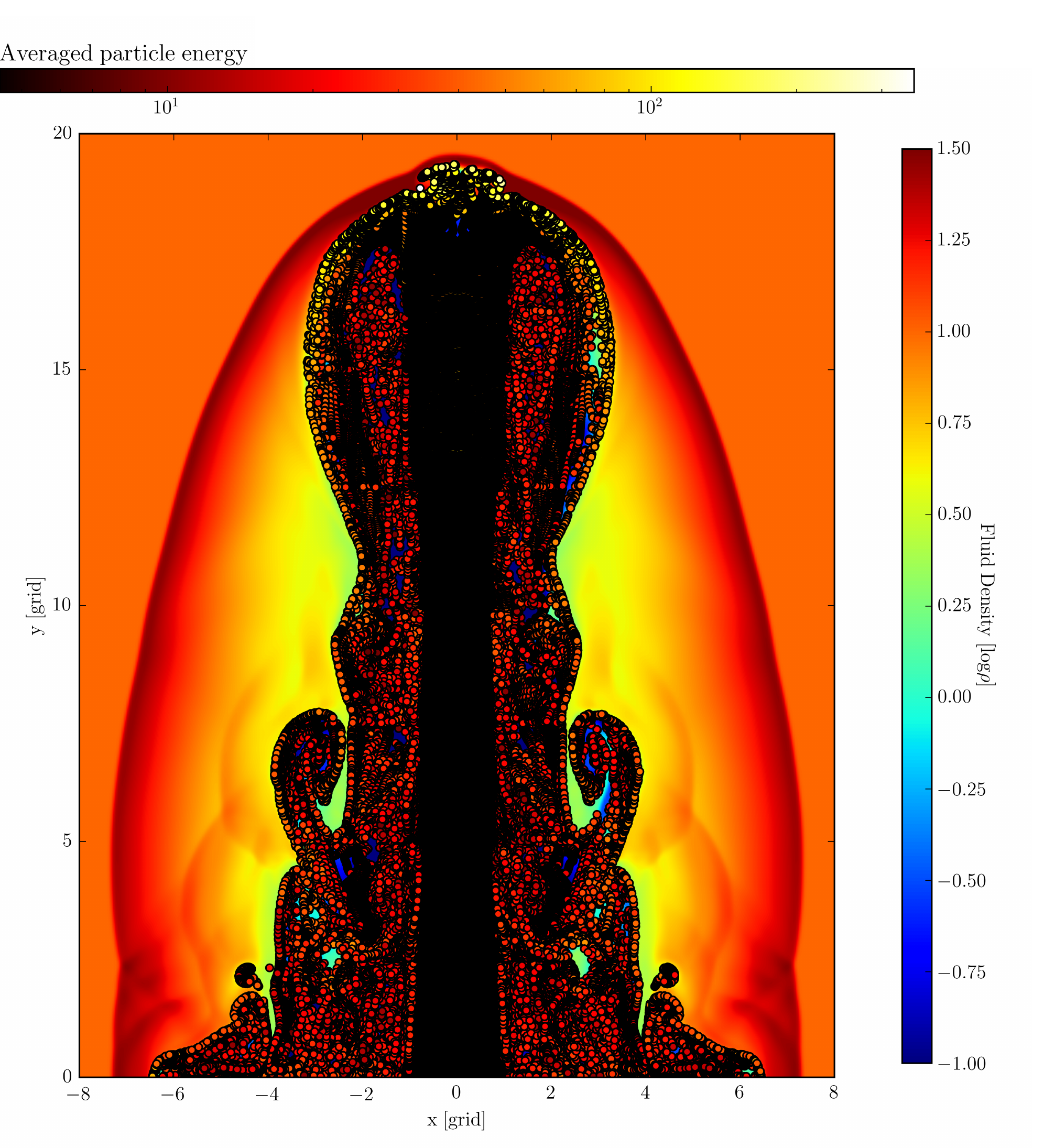}
    \caption{2D slice of a 3D hybrid fluid-particle simulation. The underlying fluid is color-coded by density in dimensionless grid units. The overplotted particle population is color-coded by averaged particle energy in dimensionless grid units, following the underlying jet stream. The figure shows simulation epoch $38$, which represents the last epoch before the jet stream moves off-grid. }
    \label{fig:epoch38}
\end{figure}

We considered a specific epoch of the jet simulation that is illustrated in
Figure~\ref{fig:epoch38}, in which the jet's hot spot or terminal shock has not yet propagated off the grid. The figure shows a 2D slice through the 3D jet simulation and (overlaid) particle population. The particle distribution is color-coded by energy on top of the thermal fluid density, both in dimensionless grid units. The particle distribution, injected into the simulation through the jet nozzle, follows the motion and speed of the underlying fluid, including the backflow motion of the jet. The particles follow the jet stream, for instance, passing through recollimation shocks. 

\begin{figure*}
        \small
        \begin{tikzpicture}
        \node[anchor=south west,inner sep=0] (Fig1) at (0,0) 
        {\includegraphics[trim={    3.1cm     0.0cm      3.3cm       0.0cm      }, clip,width=0.25\textwidth]{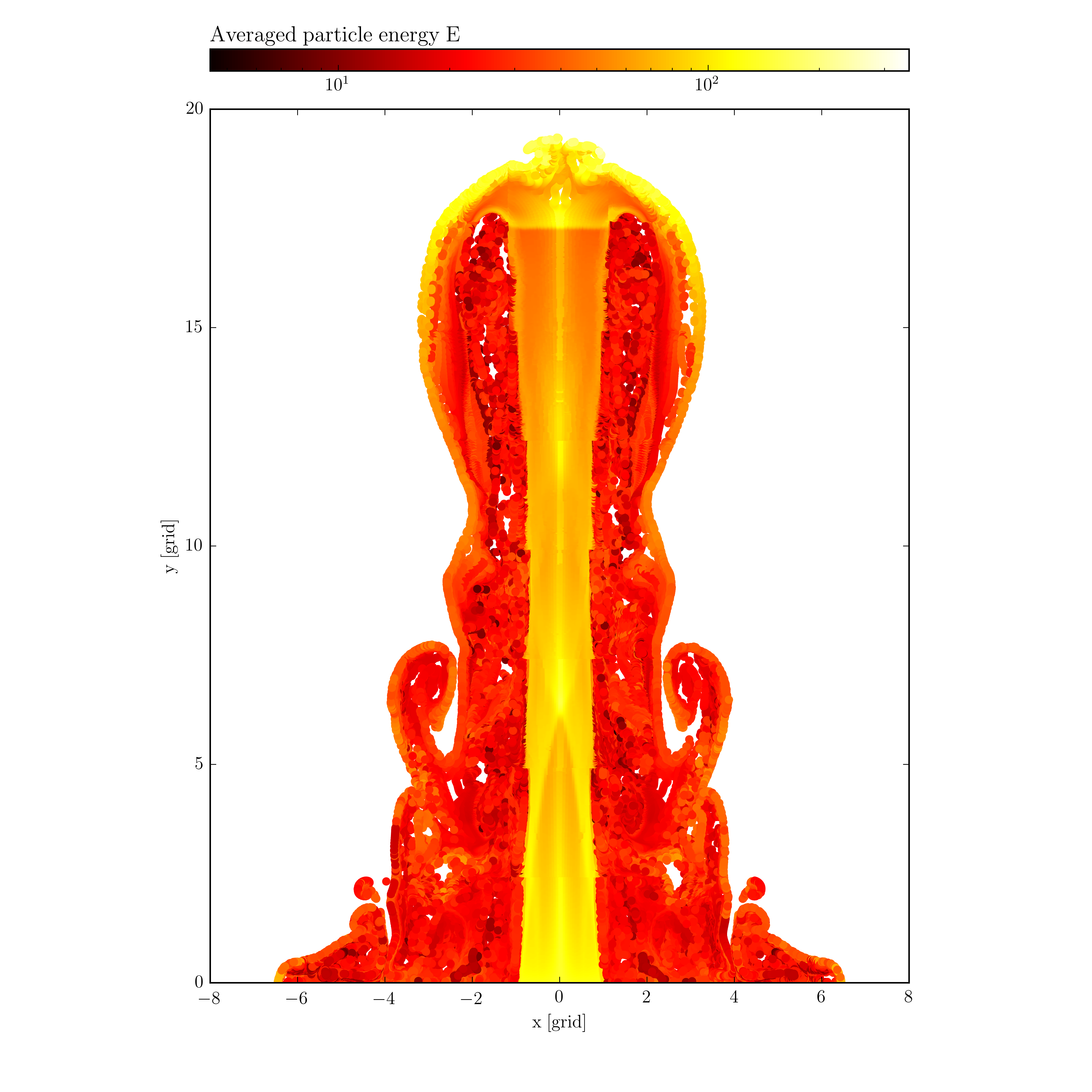}};
        \node[anchor=south west, inner sep=0] (Fig2) at (Fig1.south east) 
        {\includegraphics[trim={    3.1cm     0.0cm      3.3cm       0.0cm      }, clip,width=0.25\textwidth]{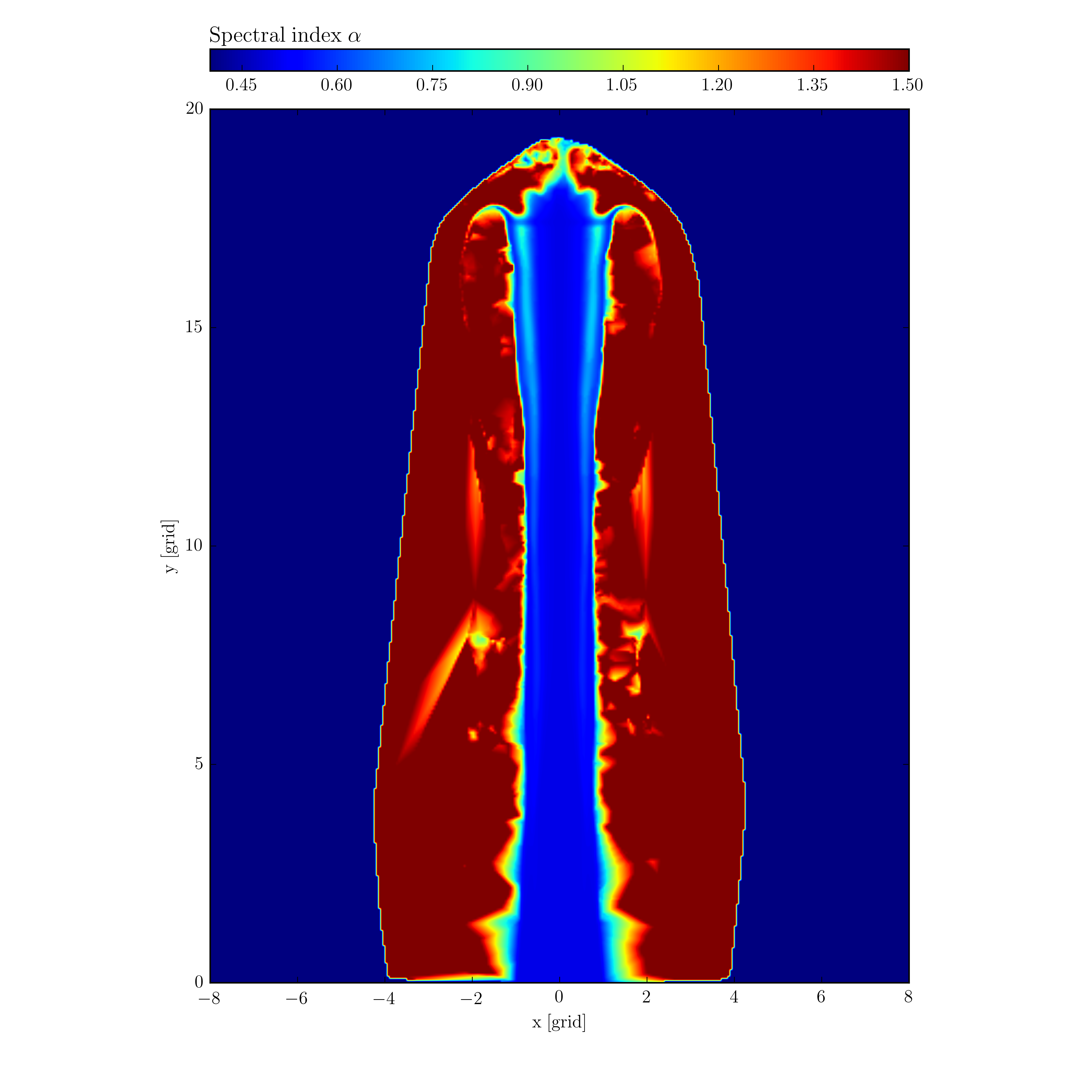}};
        \node[anchor=south west, inner sep=0] (Fig3) at (Fig2.south east) 
        {\includegraphics[trim={    3.1cm     0.0cm      3.3cm       0.0cm      }, clip,width=0.25\textwidth]{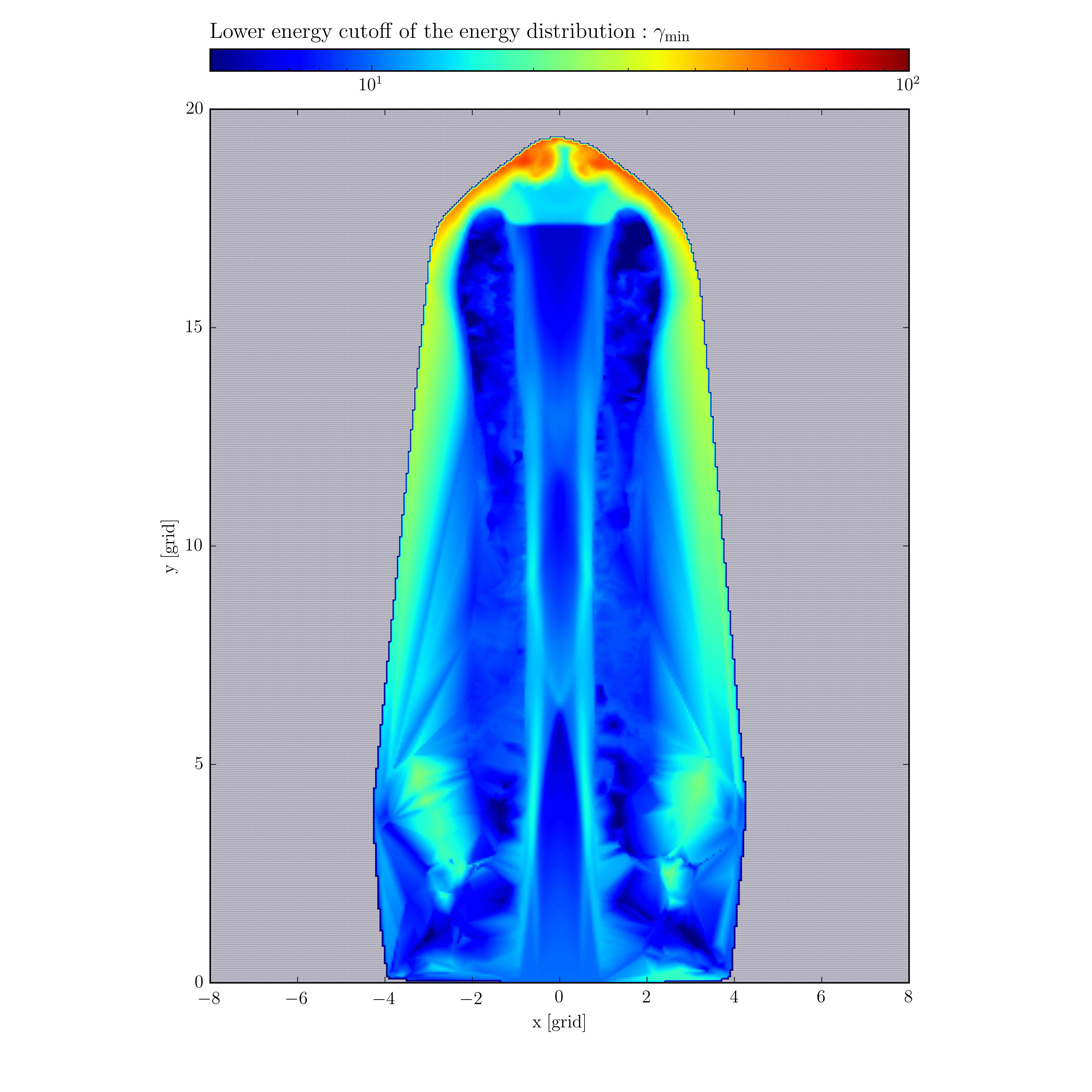}};
        \node[anchor=south west, inner sep=0] (Fig5) at (Fig3.south east) 
        {\includegraphics[trim={    3.1cm     0.0cm      3.3cm       0.0cm      }, clip,width=0.25\textwidth]{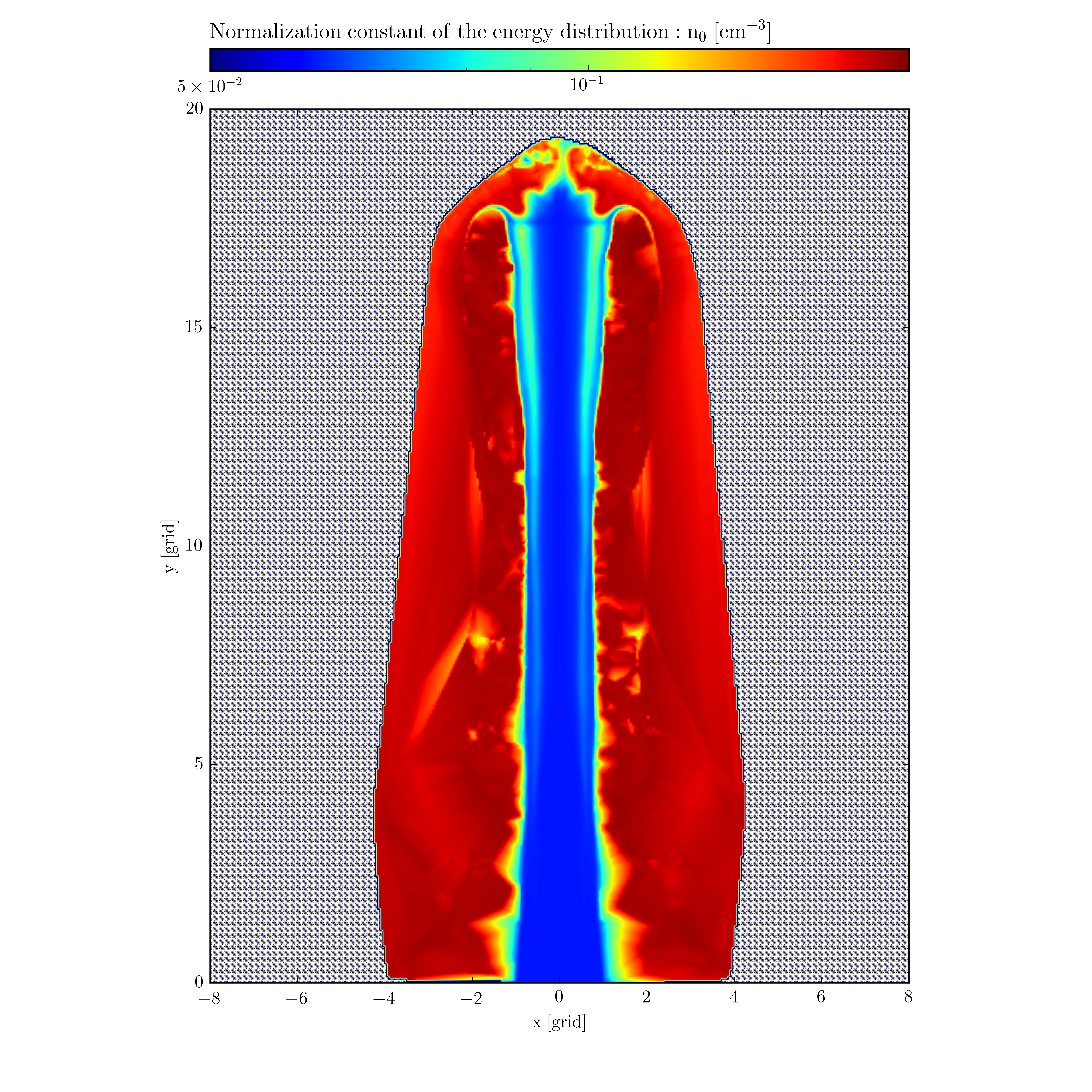}};
        \node[anchor=north,inner sep =0, label=below:$(a)$] (a) at (Fig1.south) {};
        \node[anchor=north,inner sep =0, label=below:$(b)$] (b) at (Fig2.south) {};
        \node[anchor=north,inner sep =0, label=below:$(c)$] (c) at (Fig3.south) {};
        \node[anchor=north,inner sep =0, label=below:$(d)$] (d) at (Fig5.south) {};
        \end{tikzpicture}
        \caption[]{\textbf{From left to right}: Demonstration of our synthetic imaging pipeline. All panels show a 2D slice through our 3D jet simulations. The pipeline starts to extract the particle population from the underlying fluid (see Fig.~\ref{fig:epoch38}). The particle population in panel \emph{(a)} is color-coded by dimensionless averaged particle energy. Brighter colors (white and yellow) represent higher energies, i.e., the case for the inner jet spine. The recollimation shock, associated with a centralized region of higher energy, is visible in the jet's spine. The particles are ``cooled'' by radiative losses (orange and red) and cool even more rapidly when leaving the jet flow. Panels \emph{(b)-\emph{(d)}} illustrate the interpolated particle values that form a grid of particle attributes (highest values in red): Panel \emph{(b)} shows a 3D grid of the spectral index, $\alpha,$ calculated from the spectrum's slope, $s,$ of the energetic changes due to radiative losses at a given frequency. Panel \emph{(c)} highlights the lower energy cutoff, $\gamma_\mathrm{min}$, of the particle's energy distribution. Panel \emph{(d)} shows the calculated normalization constant, $N_\mathrm{e}$, of the energy distribution interpolated onto a 3D grid. 
                }
        \label{fig:pipeline}
\end{figure*}
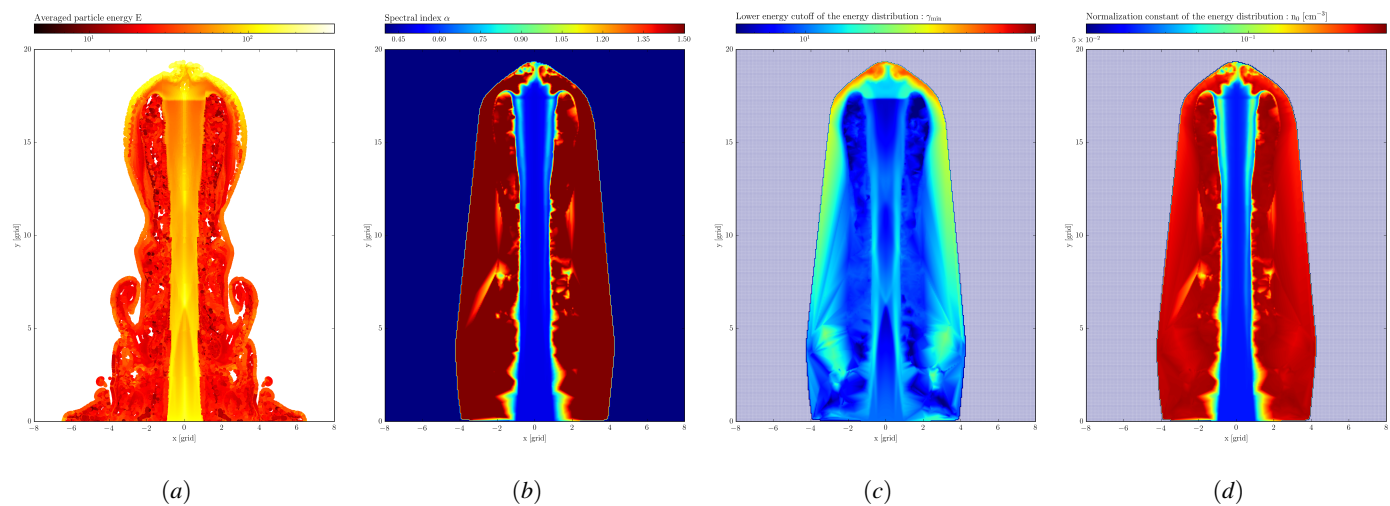
In particular, our post-process calculations are summarized (sequentially) in Fig.~\ref{fig:pipeline}:
\begin{itemize}
    \item[\textbf{1.}] Panel \emph{(a)}: The particle distribution is separated from the underlying fluid and the position of the macro-particle is stored. The distinct particles are illustrated without hard edges. This step highlights the energetic changes of the particles when leaving the jet stream. Additionally, the effect of the recollimation shock, which is the increase in the particle's energy, can be seen in this panel. 
    \item[\textbf{2.}] Panel \emph{(b)}: We interpolated the spectral index, $\alpha$ (power-law slope, $s$) from the resultant slope of the energy distribution in Fig.~\ref{fig:spectrum} at the electron energy, $\gamma,$ that corresponds to a set observational frequency, $\nu_\mathrm{obs}$. Our radiative transfer scheme relies on the calculated values for the transfer coefficients presented in ~\cite{JonesOdell}, which implies that we set a limit for the spectral index for strongly cooled particles arbitrarily to $\alpha=1.5$. With this, we subsequently defined which particles are removed from all interpolations.
    \item[\textbf{3.}] Panel \emph{(c)}: Similar to step 2, we interpolated the lower energy cutoff, $\gamma_\mathrm{min}$, that is, the lowest energy bin per particle of the simulated power-law distribution in dimensionless units. 
    \item[\textbf{4.}] Panel \emph{(d)}: In a similar manner, we computed the normalization constant, $N_\mathrm{e}$, from $\gamma_\mathrm{min}$ and the power-law index, $s$ (see Eq.~\ref{eq:norm}) of the energy distribution in Eq.~\ref{eq:powerlaw} at a desired frequency, and therefore at a specific energy. 
\end{itemize}
Finally, we output all three components of the magnetic field $B_{x,y,z}$ (as well as the strength) , the lower energy cutoff, $\gamma_\mathrm{min}$, the normalization constant, $N_\mathrm{e}$, and the spectral index, $\alpha$, on a 3D Cartesian grid into files adjusted for RADMC-3D. In all our RMHD jet simulations, each computational box consists of $320 \times 320 \times 400$ zones, and we set the viewing distance of the source at a redshift of $z=0.002$~\citep[the redshift of Centaurus A,][]{CenAredshift}. The individual scaled cell size is 0.01\,parsec (pc). For all the images presented in Sect.~\ref{sec:4}, we used an observing frequency of $\nu_\text{obs} = 86\,$GHz.

\section{Synthetic polarized emission calculations}\label{sec:3}
RADMC-3D is a thoroughly tested and well-documented ray-tracing code for solving astrophysical radiative transfer problems~\citep{Dullemond}. RADMC-3D computes the Stokes parameters $I, Q, U$, and $V$ (total intensity, LP, and CP, respectively) along individual rays. A modified version of the code accounts for synchrotron absorption, emissivity, Faraday conversion, as well as Faraday rotation~\citep{MacDonald2018}. The solution of the emitted polarized synchrotron radiation is a function of optical depth, the normalization constant, $N_\mathrm{e}$, the low-energy cutoff, $\gamma_\text{min}$, of the power-law distribution in Eq.~\ref{eq:powerlaw}, the strength of the magnetic field, and its orientation to our line of sight. 
In contrast to~\cite{Kramer&MacDonald}, we made use of the updated implementation of the full Stokes module in RADMC-3D by~\cite{Macdonald2021}. This allows for an additional degree of freedom, namely, a full rotational view of our 3D solution. Finally, we adapted the above described implementation to solve for a variable spectral index. 
In particular, our solution will apply
the radiative transfer calculations to interpolated particle attributes that connect the emitter with the observer. This involves two main steps: \emph{(i)} a relativistic aberration correction, which is applied cell-by-cell to determine the angle between the local magnetic field vector of each cell and the observer's orientation relative to the jet axis, and \emph{(ii)} a rotation correction, which is applied in the same way to convert the LP ellipse from the local comoving frame onto the plane of the sky. Further details on these corrections are published in~\cite{Macdonald2021}.\footnote{Please note that we do not apply a slow-light approach in this work.}

Once the Stokes parameters are computed for each line of sight (corresponding to each pixel/ray in our synthetic maps), a Doppler factor is introduced (see Eq.~\ref{eq:doppler}). This factor accounts for the bulk Lorentz factor and orientation of the large-scale jet structure and is used to transform the flux levels into the observer's frame.

\section{Hybrid fluid-particle jet simulation results}\label{sec:4}
\subsection{Centaurus A as a laboratory}
Centaurus A (Cen\,A) is one of the earliest known radio galaxies~\citep{Clarke}, and is located at a distance of approximately 3.8 million parsecs~\citep[$z=0.002$][]{Harris2010,CenAredshift}. It hosts a SMBH at its core with an estimated mass of about 55 million solar masses~\citep{Neumayer}. Cen\,A offers a rich field for multiwavelength observations, though this paper primarily explores its radio characteristics. Cen A features strong jets visible in both the radio and X-ray spectra, spanning from sub-parsecs to hundreds of parsecs in scale~\citep{Clarke,Hardcastle,Feain,Muller2011,muller2014,Ehlert2022}.
The jet's observed inclination ranges from $i \in [12-45]^\circ$~\citep{muller2014}, with edge-brightening supporting a spine-sheath jet structure~\citep{JanssenCenA,Laing99}. We set our simulated jet's inclination to $i=35.5^\circ$. Cen\,A is an exemplary case study for AGN research in numerical simulations, particularly for assessing the impact of synchrotron losses in relativistic jet models.

\subsection{The effect of synchrotron cooling}
\begin{figure}
    \begin{tikzpicture}
    \node[anchor=north west,inner sep=0] (Fig1) at (0,0) 
    {\includegraphics[trim={    0.5cm     0.0cm      2.1cm       0.9cm      }, clip,width=0.46\textwidth]{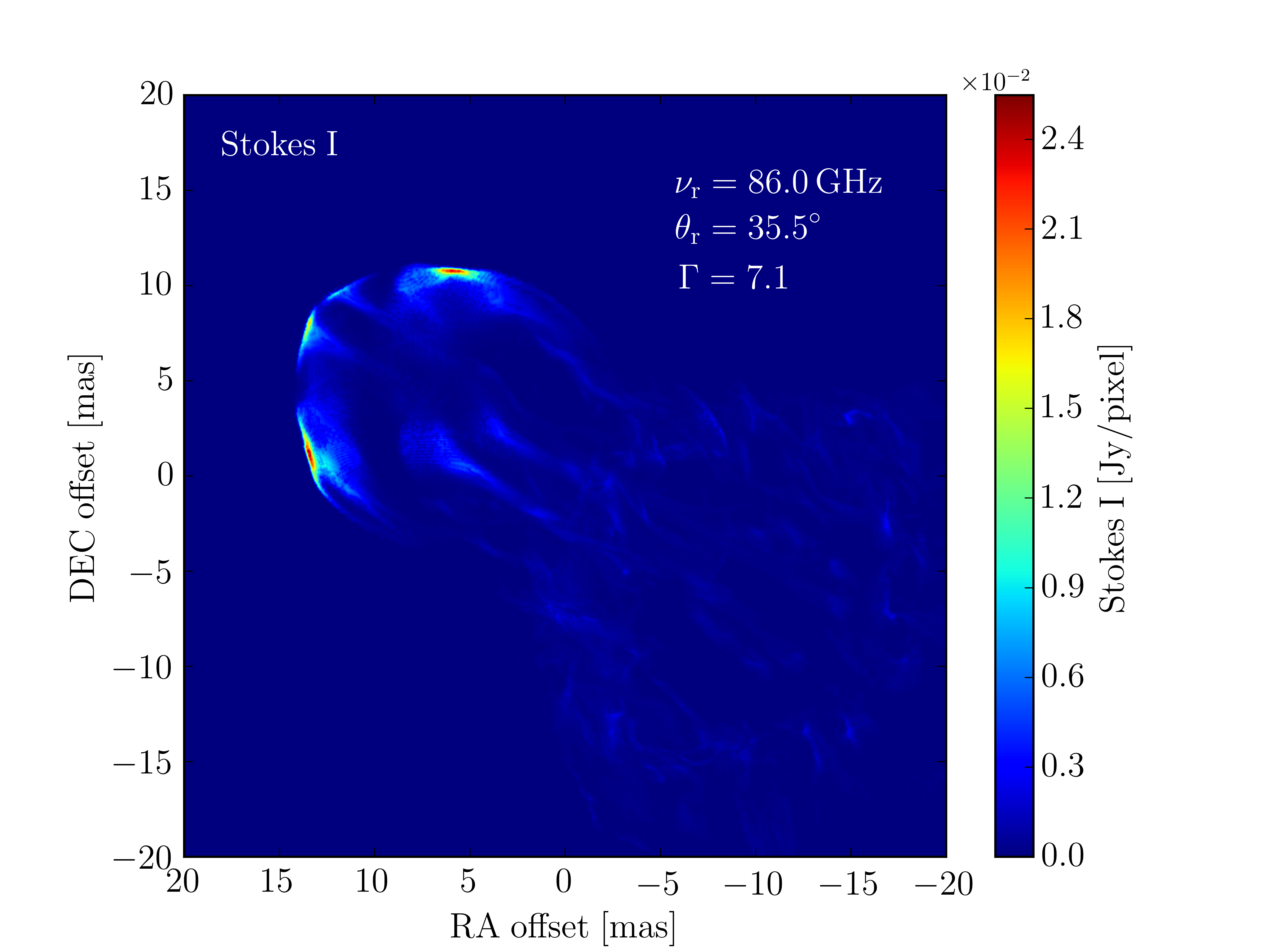}};
    \node[anchor=north west, inner sep=0] (Fig2) at (Fig1.south west) 
    {\includegraphics[trim={    0.5cm     0.0cm      2.1cm       0.8cm      }, clip,width=0.46\textwidth]{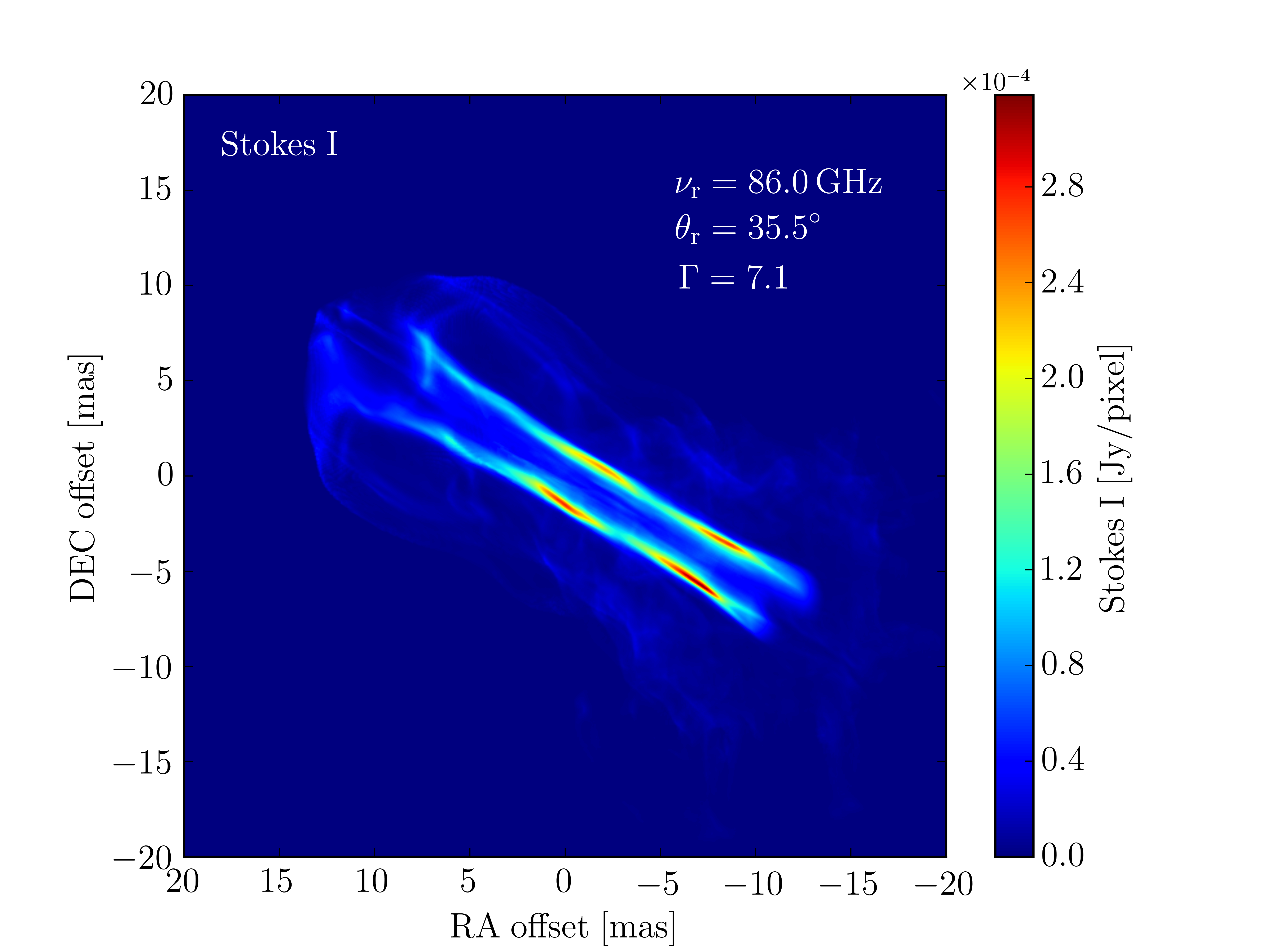}};
    \node[anchor=north west, inner sep=0] (Fig3) at (Fig2.south west) 
    {\includegraphics[trim={    0.5cm     0.0cm      2.1cm       0.8cm      }, clip,width=0.46\textwidth]{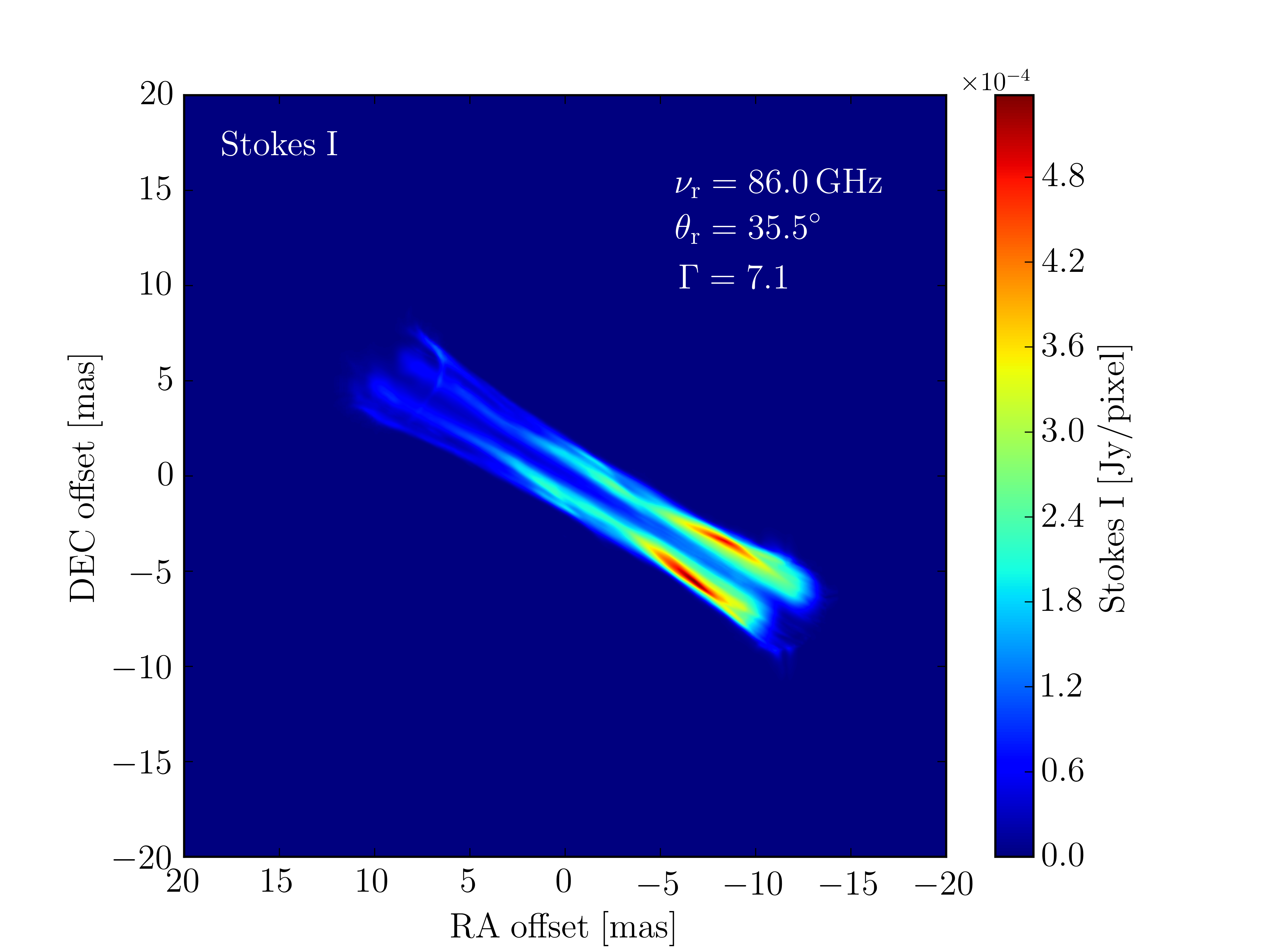}};
    \end{tikzpicture}
    \caption{Synthetic total intensity map of the 3D RMHD jet simulation. \textbf{Top}: Map of the 3D RMHD jet simulation generated without the use of a jet tracer, mainly illustrating the emission of the ambient medium. \textbf{Middle}: Map of the jet simulation generated by applying an arbitrarily chosen jet tracer to exclude the ambient medium, highlighting an edge-brightened jet. \textbf{Bottom}: Final 3D hybrid fluid-particle jet simulation, showing an edge-brightened jet emission with the ambient medium excluded by including radiative losses. All maps are ray-traced at 86\,GHz and viewed at an observational angle of 35.5$^\circ$.}
    \label{fig:comp}
\end{figure}

The PLUTO code can keep track of an optional jet tracer value for each cell in the simulated Eulerian grid, ranging from zero (in the ambient medium) to one (in the jet). A value close to one indicates that the cell belongs to the jet, while a value close to zero indicates a high presence of ambient medium. By selecting a cutoff value near unity, we can exclude the ambient medium from the input grid that RADMC-3D uses for ray-tracing when creating full Stokes emission maps. This exclusion was necessary for the non-hybrid jet simulations presented in~\cite{Kramer&MacDonald}, where we did not account for the effect of synchrotron cooling (specifically, a cutoff value of 0.75 was used; see the \textbf{middle} panel in Fig.~\ref{fig:comp})\footnote{It is noteworthy that an additional cut of the grid is applied to the data in order to exclude the jet's bulk in~\cite{Kramer&MacDonald}.}. Consequently, in these initial calculations, the ambient medium does not cool over time and emits a large amount of synchrotron emission, which obscures our view of the inner jet spine (i.e., with no tracer; see the \textbf{top} panel of Fig.~\ref{fig:comp}). An alternative view of this is illustrated in Fig.~\ref{fig:log}. As a result, the jet is entirely obscured by the ambient medium in the absence of synchrotron cooling.

In contrast, when we incorporate synchrotron cooling (via our hybrid fluid-particle simulation), we can clearly observe the jet distinctly from the ambient medium (e.g., \textbf{bottom} panel of Fig.~\ref{fig:comp}). This was achieved without using any jet tracer to arbitrarily exclude the emission of the ambient medium.
The emission calculated solely from particle attributes confirms our findings in~\cite{Kramer&MacDonald}, that the toroidal magnetic field configuration consistently results in emission concentrated along the edges of the jet. The simulations in this work produce an edge-brightened appearance in all presented synthetic synchrotron emission maps. Furthermore, the presence of both positive and negative CP, visible in the jet emission, indicates the changing orientation of the jet's magnetic field relative to our line of sight (see Fig.~\ref{fig:pol}).

\subsection{Synthetic synchrotron emission maps}
In our hybrid fluid-particle simulation, we once again investigated a toroidal magnetic field morphology within the jet. We selected the toroidal magnetic field configuration to replicate the edge-brightening effect observed in the radio galaxy Centaurus A~\citep{JanssenCenA,Kramer&MacDonald}.
In Fig.~\ref{fig:comp}, we compare the results of the non-particle approach, depicted in the top and middle panels (with and without a jet tracer, respectively), to the corresponding images obtained from the hybrid fluid-particle simulations, illustrated in the \textbf{bottom} panel. The figures display the total intensity of the jet's emission at an inclination of $i=35.5^\circ$, observed at $86\,$\,GHz (alternative frequencies are presented in Fig.~\ref{fig:app} and discussed in Appendix~\ref{app:freq}). The image clearly demonstrates that the use of a tracer is unnecessary if the simulation accounts for radiative losses. This is stressed as we observe a distinct and well-defined jet structure in the hybrid fluid-particle simulation.
We can further compare the resultant brightness (in Jansky per pixel) of the jet between the different approaches, namely, the hybrid fluid-particle approach and the pure fluid jet. The difference between the bow shock's luminosity and the rest of the relativistic jet makes a noticeable impact on the resultant brightness. When the bow shock is not excluded from the simulation, it outshines the jet structure completely. The bow shock seems to be two orders of magnitude brighter (i.e., $\sim 2.4 \times 10^{-2}\,$Jy/pixel) than the jet structure ($\sim 2.8\times 10^{-4}\,$Jy/pixel). The brightness and structure of the hybrid-fluid particle simulation and the RMHD jet simulation (using a jet tracer to exclude the lobe) seem to be in agreement. The relativistic jet is carrying an underlying toroidal magnetic field in all panels depicted in Fig.~\ref{fig:comp}. The resultant structure resembles the results published in~\cite{Kramer&MacDonald}. When the jet is visible, an edge brightened jet structure is observed. Both the hybrid fluid-particle jet and the pure fluid jet show the behavior of maximum luminosity in both edges close to the jet launching in the lower right part of the figures. The hybrid fluid-particle jet excludes more of the upstream-structure of the jet due to opacity effects incorporated into our calculations. 

\begin{figure}
    \begin{tikzpicture}
    \node[anchor=north west,inner sep=0] (Fig1) at (0,0) 
    {
    \includegraphics[trim={    0.5cm     0.0cm      2.1cm       0.9cm      }, clip,width=0.46\textwidth]{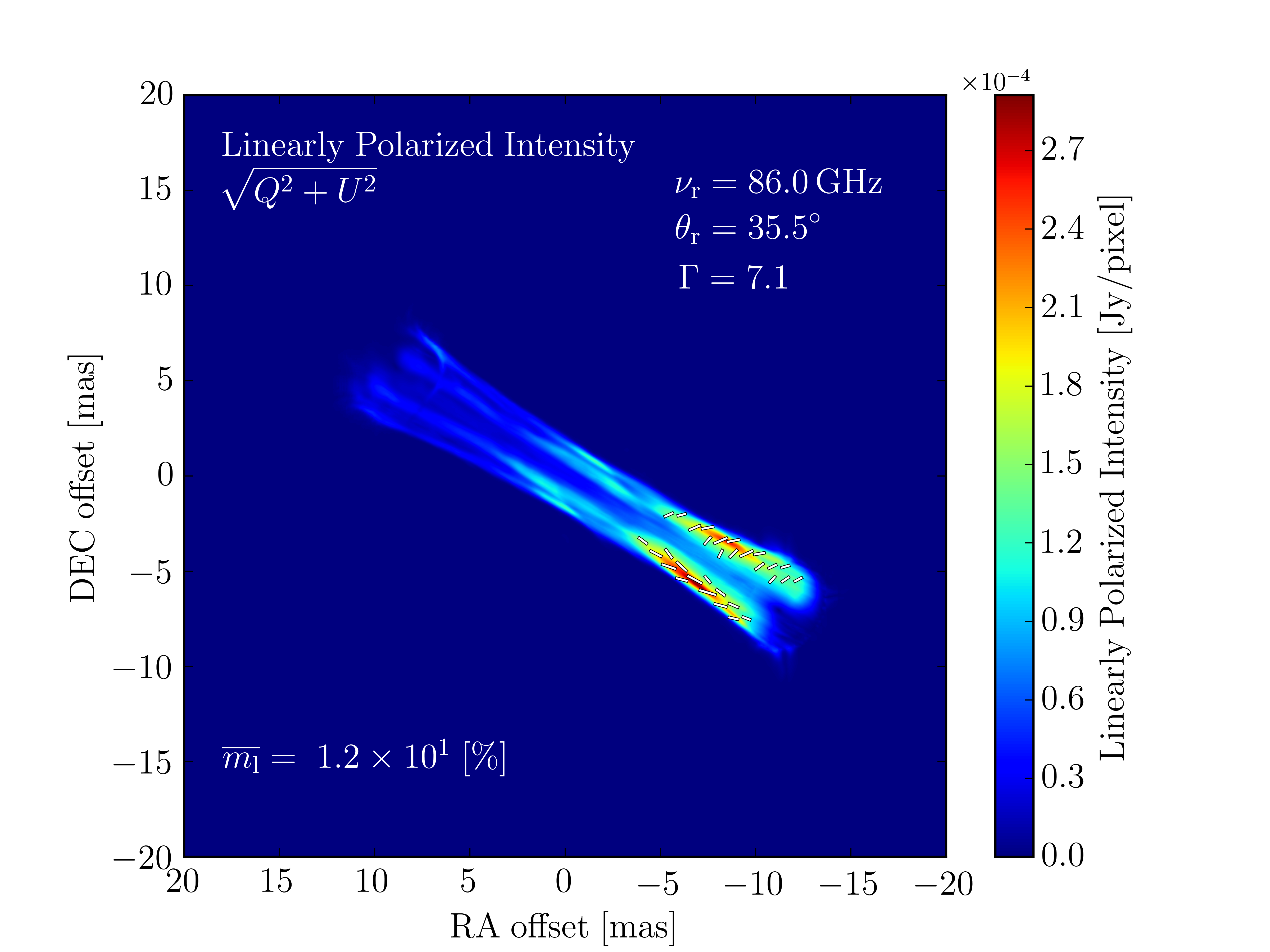}};
    \node[anchor=north west, inner sep=0] (Fig2) at (Fig1.south west) 
    {
    \includegraphics[trim={    0.6cm     0.0cm      1.75cm       0.8cm      }, clip,width=0.46\textwidth]{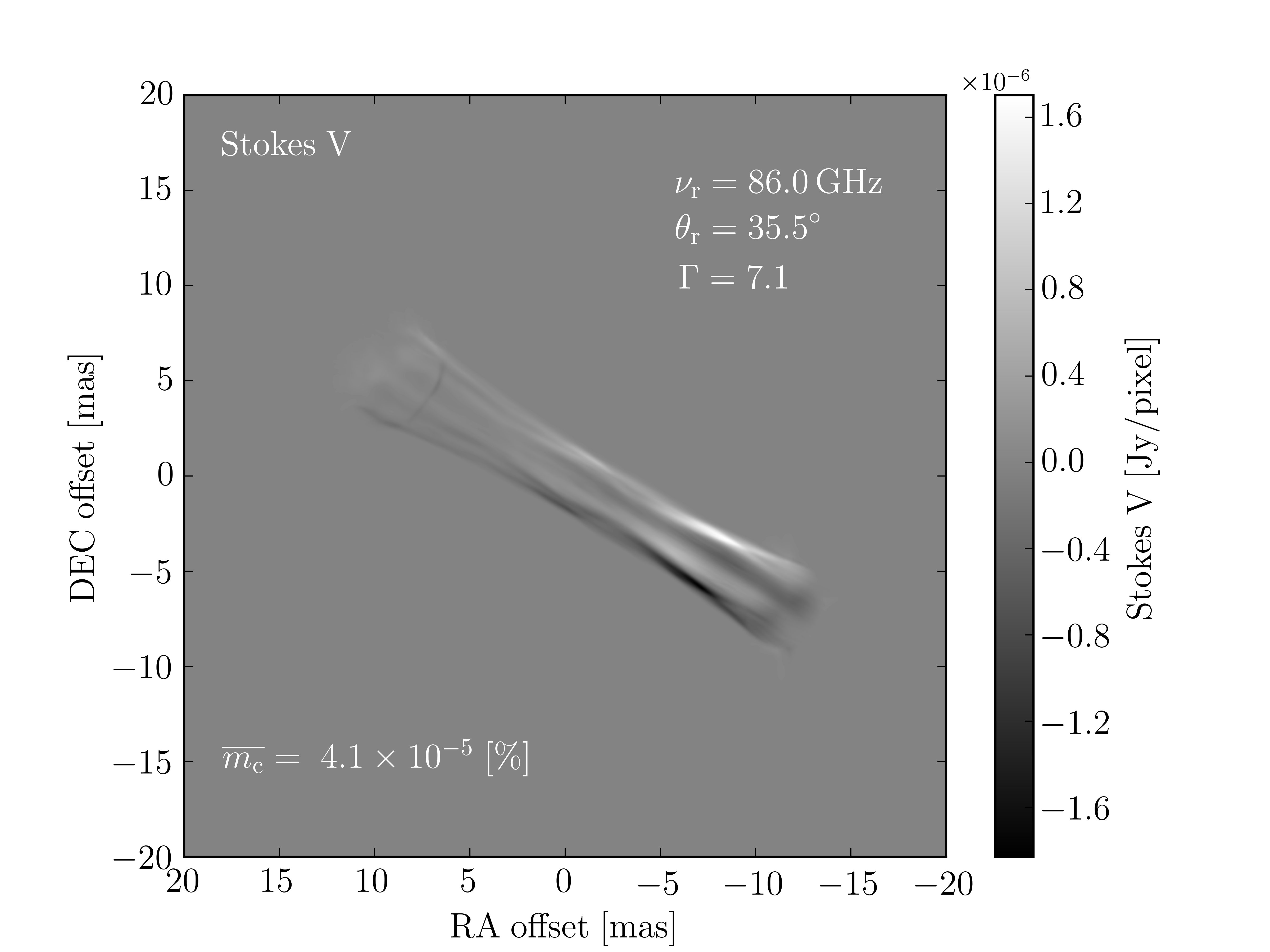}};
    \end{tikzpicture}
     \caption{Ray-traced images of our hybrid fluid-particle jet propagating from the bottom right to the top left in linearly polarized intensity (\textbf{top}) and CP (\textbf{bottom}). The pictures illustrate the jet viewed at an observational viewing angle of $35.5^\circ$ at 86\,GHz. The fractional linear ($m_\mathrm{l}$) and circular ($m_\mathrm{c}$) polarization level is given in the bottom-left corner of each map.}
    \label{fig:pol}
\end{figure}

In Fig.~\ref{fig:pol}, we present images of the associated synchrotron polarization. The \textbf{top} panel displays the linearly polarized intensity, which is calculated as the square root of the sum of the squares of the Stokes parameters $Q$ and $U$. It also includes the electric vector position angles (EVPAs) computed as $\chi= 0.5 \arctan(U/Q)$. The EVPA convey the orientation of the magnetic field. As the viewing angle is $i=35.5^\circ$, the jet is partially boosted to or de-boosted away from our line of sight. The lower/left edge of the jet carries EVPAs oriented alongside the jet's direction, highlighting a magnetic field that takes a turn around that edge. The upper, right edge of the jet shows EVPAs perpendicular to the jet's direction in the innermost part and a small rotation toward the edges. This is indicative of a magnetic field orientation along the jet before taking another turn around the edges. The \textbf{bottom} panel shows the CP, namely, Stokes $V$. Similar to \cite{Kramer&MacDonald}, we observe a flip in sign of CP with one sign on each of the highlighted edges. The lower edge carries a strong negative Stokes $V$, aligning with EVPAs oriented in the jet direction, whereas the upper edge is positive in Stokes $V$. This again conveys the underlying toroidal magnetic field configuration.

We computed the integrated levels of fractional polarization, which are flux-weighted averages of the Stokes parameters across the entire jet emission region in each set of images~\citep[we followed the formalism presented in][]{Kim2019}. These values are listed to the lower left in the linearly polarized and circularly polarized images of Fig.~\ref{fig:pol}. The values are denoted as $\bar{m_l} \equiv \sqrt{\bar{Q}^2 + \bar{U}^2}/\bar{I}$ for LP and $\bar{m_c} \equiv -\bar{V}/\bar{I}$ for CP. The fractional LP is $m_\mathrm{l}=1.2\times10^1\,\%$, while the fractional CP is $m_\mathrm{c}=4.1\times10^{-5}\,\%$. In circularly polarized synchrotron emission, we observe both positive and negative CP, which emphasizes the dynamic changes in the orientation of the jet's magnetic field relative to our line of sight. This finding in our hybrid fluid-particle jet simulation agrees with the RMHD jet simulation presented in~\cite{Kramer&MacDonald}.

\section{Discussion}\label{sec:5}
Our spectral analysis of the particles' energy over time confirms that the particles in the relativistic jet experience synchrotron losses (see Fig.~\ref{fig:spectrum}). Notably, higher-energy electrons within the macro-particle exhibit a more rapid cooling rate compared to their lower-energy counterparts. Consequently, the initial spectrum's high-energy tail experiences decay, leading to a subsequent reduction in the parameter $\gamma_\mathrm{max}$~\citep[This finding is confirmed in ][]{Dubey2023}.

Furthermore, particles that are impacted by the jet's recollimation shock are gaining energy (see Fig.~\ref{fig:pipeline} second image from left). A comprehensive study on the particle shock acceleration is presented in, for example, \cite{Mukherjee21}, \cite{Borse}, \cite{Girib}, and \cite{Dubey2023}. 

We compared our synthetic polarized emission maps with observational data, focusing on the TANAMI~\citep{Ojha2010} observations by \cite{muller2014} and the Event Horizon Telescope (EHT) observations by \cite{JanssenCenA}. 
The interpretation of the brightest jet features as the radio core is favored by~\cite{JanssenCenA}, implying the presence of a stationary shock feature. They consider this a likely explanation for their EHT observations, given the small scales captured in the image presented in~\cite{JanssenCenA}, the high observation frequency, and the closeness of the source. However, they do not eliminate the prospect of insufficient particle acceleration being the main driver for the brightest jet features.

\cite{muller2014} discovered a prominently bright jet feature, J$_\text{stat}$, situated approximately 3.5 milliarcseconds (mas) downstream from the core, observed to remain stationary relative to the core. This feature's position could possibly be identified with our simulation's brightest feature, potentially aligning it with either the radio core observed by \cite{muller2014} or their J$_\text{stat}$ component. If our simulated jet nozzle corresponds to the radio core, the 3.5\,mas distance would accurately represent J$_\text{stat}$. \cite{Tingay} also observed this component, referred to as C3. J$_\text{stat}$ could be interpreted as a ``cross-shock''\footnote{Cross-shock features are expected to exhibit a spectral inversion linked to recollimation shocks, where components passing through are accelerated in subsequent expansions~\citep{muller2014}.} in the jet flow, similar to phenomena observed in simulations of over-pressured jets~\citep{Mimica2009}, where a spectral inversion occurs at recollimation shocks, leading to the acceleration of components in subsequent expansions.
Although our simulations do show a recollimation shock in the underlying RMHD simulation, our interpretation diverges. We attribute this phenomenon mainly to the accumulation and amplification of synchrotron emission around the jet edges, driven by the presence of a toroidal magnetic field component. Notably, our numerical model reproduces the observed emission without requiring the introduction of any strong shock acceleration. We stress that a purely toroidal magnetic field configuration would yield symmetric edge-brightened emission spanning the jet, as discussed in~\cite{Gabuzda2018}. Conversely, a helical magnetic field configuration would lead to more asymmetric emission along the jet edges. This is confirmed in numerical simulations by \cite{Kramer&MacDonald}, and in the work presented in this paper. We validated the edge-brightened structure of an underlying toroidal magnetic field configuration when calculating the synthetic polarized synchrotron emission from NTP attributes directly. We, moreover, infer that the intrinsic magnetic field morphology is toroidal in nature due to the edge-brightened profile observed by \cite{muller2014}  and \cite{JanssenCenA}. 

Our perspective aligns with the interpretation on the total intensity outlined in \cite{JanssenCenA}, as we also observe a ``core'' shift in the peak of Stokes $I$ and linearly polarized intensity as we transition from lower to higher frequencies (from 15\,GHz to 230\,GHz; see Fig.~\ref{fig:app}). We observe that the jet appears optically thick upstream, while downstream, it becomes optically thin. \cite{JanssenCenA} identifies the most prominent features within the jet as radio cores, signifying the boundary between regions of upstream synchrotron self-absorption and downstream optically thin regions. They managed to resolve the self-absorbed section between the proposed radio core and the jet apex, a location that coincides with the presence of the SMBH and its accretion disk.\footnote{It is important to acknowledge that with existing telescopes, the radio core and upstream region typically remain unresolved for the majority of AGN.} The interpretation of the brightest jet feature as a radio core, given their data, indeed appears to be the most plausible explanation. According to jet theory, a luminous radio core should be observable in very long baseline interferometry images. This phenomenon is a common occurrence in sources resembling Centaurus A, and the core shift usually adheres to the standard $\nu^{-1}$ relation, consistent with most sources~\citep{89}. 

Furthermore, our spectral index cube in Fig.~\ref{fig:pipeline} (b), which interpolates particle attributes across a 3D domain, alongside the jet's edge-brightening, suggests a transition in opacity from the inner relativistic jet flow to the surrounding backflow material. This phenomenon aligns with the concept of a spine-sheath jet structure~\citep[e.g.,][]{Laing99,Ghisellini,muller2014}, where a faster-moving spine becomes visible due to its smaller optical depth. \cite{Worrall2008} also noted a steeper spectrum in the jet's outer layers, reinforcing the idea of a spine-sheath configuration for jets at the 100\,pc scale.

Overall, \cite{muller2014} observed  jet emission extending up to approximately 70\,mas, or about 1.3\,pc, displaying a straight and well-collimated shape without significant bends. This is confirmed by our simulations, revealing a stable, collimated outflow up to several parsec away from the jet launching (nozzle) region.

Our simulations indicate that Cen A presents the potential to reveal a polarized signal up to $\bar{m}_l=12\,\%$ (Fig.~\ref{fig:pol}) in fractional linear polarization and $\bar{m}_c=4.1\times10^{-5}\,\%$ in fractional CP within its compact jet region, information that is currently unavailable in observations. Acquiring these data is crucial for investigating the behavior of the magnetic field on the inferred scales and assessing the validity of the strong toroidal component generating a bi-modal EVPA pattern and a switch in sign of circularly polarized synchrotron emission. Furthermore, if beam depolarization is responsible for the non-detection of polarization observed by the Atacama Large Millimeter Array (ALMA), conducting observations at 43\,GHz wavelength might offer one of the most promising opportunities for polarization detection. This is due to the anticipated detection of the bright jet sheath with a relatively small beam size. 86\,GHz observations could serve as calibrator and to validate a possible outcome of nonzero polarization in Centaurus A. 

\section{Conclusion}\label{sec:6}
We used the PLUTO code~\citep{Mignone2007} to compute 3D RMHD simulations with a Lagrangian particle module~\citep{Vaidya2018}.
An analysis of the energy distribution of particles over time reveals the impact of synchrotron losses on both the particles within the ambient medium and those within the relativistic jet. To generate ray-traced full Stokes images from these fluid-particle hybrid jet simulations, we performed an interpolation of discrete particle energy and particle number density values across particle positions. Subsequently, these interpolated values were mapped onto continuous Cartesian grids, which were further employed as inputs for the ray-tracing code RADMC-3D. This process allowed us to construct a variable spectral index cube. Importantly, the hybrid fluid-particle jet methodology enables the generation of synthetic emission maps that emphasize the jet's (polarized) synchrotron emission, obviating the need for the arbitrary ``jet tracer'' commonly employed in conventional approaches to exclude the ambient medium. This is particularly significant as, in the context of 3D relativistic MHD simulations, the presence of the ambient medium, in the absence of synchrotron losses, can obscure the jet.

The conclusions drawn from our study can be summarized as follows:
\begin{itemize}
\item We calculated the synthetic polarized synchrotron emission directly from NTP attributes and verify the presence of the edge-brightened structure, which corresponds to an underlying toroidal magnetic field configuration~\citep{Kramer&MacDonald}.
\item Our calculations suggest that the edge-brightened jet structure apparent in EHT observations of Centaurus\,A~\citep{JanssenCenA} can be attributed to the presence of a large-scale helical magnetic field within the jet. A dominant toroidal field component is favored. This is achieved without the need to introduce any substantial shock acceleration (see, e.g., Fig.~\ref{fig:app}).
\item From the nearly symmetric brightness profile observed in the data by~\cite{JanssenCenA}, we infer that the intrinsic magnetic field morphology in Centaurus A is most likely toroidal in nature.
\item Our observations reveal a variation in the total intensity peak as we move from lower to higher frequencies, indicating a ``core'' shift. Meanwhile, we also note that the jet is optically thick upstream but becomes optically thin downstream (see Fig.~\ref{fig:app}).
\end{itemize}

Future refinements will involve: adjusting the Lorentz factor to better match the characteristics of Centaurus A; incorporating shock acceleration to improve the accuracy of the spectral index map across the grid (although the difference is not major, the toroidal magnetic field configuration lacks a recollimation shock); making dimensional adjustments to align our polarized maps with the field of view in the EHT observations; and proposing new observations at 86\,GHz of Centaurus\,A to complement and enhance our study.

\vspace{20pt}
\begin{acknowledgements}
        J. A. K.'s research was supported through a PhD grant from the International Max Planck Research School (IMPRS) for Astronomy and Astrophysics at the Universities of Bonn and Cologne, Germany. She is supported for her research by a NASA/ATP project. The LA-UR number is LA-UR-24-22728.
        The 3D jet simulations presented in this paper have been performed using the PLUTO Code. The calculations were performed on the Raven high performance cluster at the Max Planck Computing and Data Facility. The ray-tracing software RADMC-3D was used to generate the polarized images of the synchrotron emission. The authors thank B. Vaidya and D. Mukherjee for important and thoughtful discussions on the results presented in this work. 
        The authors are grateful to E. Ros for comments and M. Janssen for insights on Centaurus\,A. 
        G. F. P. acknowledges support by the European Research Council advanced grant “M2FINDERS - Mapping Magnetic Fields with INterferometry Down to Event hoRizon Scales” (Grant No. 101018682).
        L. R. is funded by the Deutsche Forschungsgemeinschaft (DFG, German Research Foundation) – project number 443220636. 
        
\end{acknowledgements}
\bibliographystyle{aa}
\bibliography{thesis_refs}

\begin{thebibliography}{80}
\expandafter\ifx\csname natexlab\endcsname\relax\def\natexlab#1{#1}\fi

\bibitem[{{Bai} {et~al.}(2015){Bai}, {Caprioli}, {Sironi}, \& {Spitkovsky}}]{Bai}
{Bai}, X.-N., {Caprioli}, D., {Sironi}, L., \& {Spitkovsky}, A. 2015, \apj, 809, 55

\bibitem[{{Bellan}(2018)}]{Bellan}
{Bellan}, P.~M. 2018, Physics of Plasmas, 25, 055601

\bibitem[{{Blandford} \& {Rees}(1974)}]{BlandfordRees}
{Blandford}, R.~D. \& {Rees}, M.~J. 1974, \mnras, 169, 395

\bibitem[{{Borse} {et~al.}(2021){Borse}, {Acharya}, {Vaidya}, {Mukherjee}, {Bodo}, {Rossi}, \& {Mignone}}]{Borse}
{Borse}, N., {Acharya}, S., {Vaidya}, B., {et~al.} 2021, \aap, 649, A150

\bibitem[{{Brunetti} {et~al.}(2003){Brunetti}, {Mack}, {Prieto}, \& {Varano}}]{Brunetti}
{Brunetti}, G., {Mack}, K.~H., {Prieto}, M.~A., \& {Varano}, S. 2003, \mnras, 345, L40

\bibitem[{{Carilli} {et~al.}(1999){Carilli}, {Kurk}, {van der Werf}, {Perley}, \& {Miley}}]{Carilli}
{Carilli}, C.~L., {Kurk}, J.~D., {van der Werf}, P.~P., {Perley}, R.~A., \& {Miley}, G.~K. 1999, \aj, 118, 2581

\bibitem[{{Clarke} {et~al.}(1992){Clarke}, {Burns}, \& {Norman}}]{Clarke}
{Clarke}, D.~A., {Burns}, J.~O., \& {Norman}, M.~L. 1992, Astrophysical Journal, 395, 444

\bibitem[{{Condon} \& {Ransom}(2016)}]{nrao}
{Condon}, J.~J. \& {Ransom}, S.~M. 2016, {Essential Radio Astronomy}

\bibitem[{{Crew} {et~al.}(2023){Crew}, {Goddi}, {Matthews}, {Rottmann}, {Saez}, \& {Mart{\'\i}-Vidal}}]{Crew}
{Crew}, G.~B., {Goddi}, C., {Matthews}, L.~D., {et~al.} 2023, \pasp, 135, 025002

\bibitem[{{Daldorff} {et~al.}(2014){Daldorff}, {T{\'o}th}, {Gombosi}, {Lapenta}, {Amaya}, {Markidis}, \& {Brackbill}}]{Daldorff}
{Daldorff}, L. K.~S., {T{\'o}th}, G., {Gombosi}, T.~I., {et~al.} 2014, Journal of Computational Physics, 268, 236

\bibitem[{{de la Cita} {et~al.}(2016){de la Cita}, {Bosch-Ramon}, {Paredes-Fortuny}, {Khangulyan}, \& {Perucho}}]{delaCita}
{de la Cita}, V.~M., {Bosch-Ramon}, V., {Paredes-Fortuny}, X., {Khangulyan}, D., \& {Perucho}, M. 2016, \aap, 591, A15

\bibitem[{Dubey {et~al.}(2023)Dubey, Fendt, \& Vaidya}]{Dubey2023}
Dubey, R.~P., Fendt, C., \& Vaidya, B. 2023, The Astrophysical Journal, 952, 1

\bibitem[{{Dullemond} {et~al.}(2012){Dullemond}, {Juhasz}, {Pohl}, {Sereshti}, {Shetty}, {Peters}, {Commercon}, \& {Flock}}]{Dullemond}
{Dullemond}, C.~P., {Juhasz}, A., {Pohl}, A., {et~al.} 2012, {RADMC-3D: A multi-purpose radiative transfer tool}, Astrophysics Source Code Library, record ascl:1202.015

\bibitem[{{Ehlert} {et~al.}(2022){Ehlert}, {Ferrazzoli}, {Marinucci}, {Marshall}, {Middei}, {Pacciani}, {Perri}, {Petrucci}, {Puccetti}, {Barnouin}, {Bianchi}, {Liodakis}, {Madejski}, {Marin}, {Marscher}, {Matt}, {Poutanen}, {Wu}, {Agudo}, {Antonelli}, {Bachetti}, {Baldini}, {Baumgartner}, {Bellazzini}, {Bongiorno}, {Bonino}, {Brez}, {Bucciantini}, {Capitanio}, {Castellano}, {Cavazzuti}, {Ciprini}, {Costa}, {De Rosa}, {Del Monte}, {Di Gesu}, {Di Lalla}, {Di Marco}, {Donnarumma}, {Doroshenko}, {Dov{\v{c}}iak}, {Enoto}, {Evangelista}, {Fabiani}, {Garcia}, {Gunji}, {Hayashida}, {Heyl}, {Iwakiri}, {Jorstad}, {Karas}, {Kitaguchi}, {Kolodziejczak}, {Krawczynski}, {La Monaca}, {Latronico}, {Maldera}, {Manfreda}, {Massaro}, {Mitsuishi}, {Mizuno}, {Muleri}, {Negro}, {Ng}, {O'Dell}, {Omodei}, {Oppedisano}, {Papitto}, {Pavlov}, {Peirson}, {Pesce-Rollins}, {Pilia}, {Possenti}, {Ramsey}, {Rankin}, {Ratheesh}, {Romani}, {Sgr{\`o}}, {Slane}, {Soffitta}, {Spandre}, {Tamagawa}, {Tavecchio}, {Taverna}, {Tawara}, {Tennant},
  {Thomas}, {Tombesi}, {Trois}, {Tsygankov}, {Turolla}, {Vink}, {Weisskopf}, {Xie}, {Zane}, {IXPE Collaboration}, {Rodi}, {Jourdain}, \& {Roques}}]{Ehlert2022}
{Ehlert}, S.~R., {Ferrazzoli}, R., {Marinucci}, A., {et~al.} 2022, Astrophysical Journal, 935, 116

\bibitem[{{EHT MWL Science Working Group} {et~al.}(2021){EHT MWL Science Working Group}, {Algaba}, {Anczarski}, {Asada}, {Balokovi{\'c}}, {Chandra}, {Cui}, {Falcone}, {Giroletti}, {Goddi}, {Hada}, {Haggard}, {Jorstad}, {Kaur}, {Kawashima}, {Keating}, {Kim}, {Kino}, {Komossa}, {Kravchenko}, {Krichbaum}, {Lee}, {Lu}, {Lucchini}, {Markoff}, {Neilsen}, {Nowak}, {Park}, {Principe}, {Ramakrishnan}, {Reynolds}, {Sasada}, {Savchenko}, {Williamson}, {Event Horizon Telescope Collaboration}, {Akiyama}, {Alberdi}, {Alef}, {Anantua}, {Azulay}, {Baczko}, {Ball}, {Barrett}, {Bintley}, {Benson}, {Blackburn}, {Blundell}, {Boland}, {Bouman}, {Bower}, {Boyce}, {Bremer}, {Brinkerink}, {Brissenden}, {Britzen}, {Broderick}, {Broguiere}, {Bronzwaer}, {Byun}, {Carlstrom}, {Chael}, {Chan}, {Chatterjee}, {Chatterjee}, {Chen}, {Chen}, {Chesler}, {Cho}, {Christian}, {Conway}, {Cordes}, {Crawford}, {Crew}, {Cruz-Osorio}, {Davelaar}, {de Laurentis}, {Deane}, {Dempsey}, {Desvignes}, {Dexter}, {Doeleman}, {Eatough}, {Falcke}, {Farah},
  {Fish}, {Fomalont}, {Ford}, {Fraga-Encinas}, {Friberg}, {Fromm}, {Fuentes}, {Galison}, {Gammie}, {Garc{\'\i}a}, {Gentaz}, {Georgiev}, {Gold}, {G{\'o}mez}, {G{\'o}mez-Ruiz}, {Gu}, {Gurwell}, {Hecht}, {Hesper}, {Ho}, {Ho}, {Honma}, {Huang}, {Huang}, {Hughes}, {Ikeda}, {Inoue}, {Issaoun}, {James}, {Jannuzi}, {Janssen}, {Jeter}, {Jiang}, {Jim{\'e}nez-Rosales}, {Johnson}, {Jung}, {Karami}, {Karuppusamy}, {Kettenis}, {Kim}, {Kim}, {Kim}, {Koay}, {Kofuji}, {Koch}, {Koyama}, {Kramer}, {Kramer}, {Kuo}, {Lauer}, {Levis}, {Li}, {Li}, {Lindqvist}, {Lico}, {Lindahl}, {Liu}, {Liu}, {Liuzzo}, {Lo}, {Lobanov}, {Loinard}, {Lonsdale}, {MacDonald}, {Mao}, {Marchili}, {Marrone}, {Marscher}, {Mart{\'\i}-Vidal}, {Matsushita}, {Matthews}, {Medeiros}, {Menten}, {Mizuno}, {Mizuno}, {Moran}, {Moriyama}, {Moscibrodzka}, {M{\"u}ller}, {Musoke}, {Mej{\'\i}as}, {Nagai}, {Nagar}, {Nakamura}, {Narayan}, {Narayanan}, {Natarajan}, {Nathanail}, {Neri}, {Ni}, {Noutsos}, {Okino}, {Olivares}, {Ortiz-Le{\'o}n}, {Oyama}, {{\"O}zel}, {Palumbo},
  {Patel}, {Pen}, {Pesce}, {Pi{\'e}tu}, {Plambeck}, {Popstefanija}, {Porth}, {P{\"o}tzl}, {Prather}, {Preciado-L{\'o}pez}, {Psaltis}, {Pu}, {Rao}, {Rawlings}, {Raymond}, {Rezzolla}, {Ricarte}, {Ripperda}, {Roelofs}, {Rogers}, {Ros}, {Rose}, {Roshanineshat}, {Rottmann}, {Roy}, {Ruszczyk}, {Rygl}, {S{\'a}nchez}, {S{\'a}nchez-Arguelles}, {Savolainen}, {Schloerb}, {Schuster}, {Shao}, {Shen}, {Small}, {Sohn}, {Soohoo}, {Sun}, {Tazaki}, {Tetarenko}, {Tiede}, {Tilanus}, {Titus}, {Toma}, {Torne}, {Trent}, {Traianou}, {Trippe}, {van Bemmel}, {van Langevelde}, {van Rossum}, {Wagner}, {Ward-Thompson}, {Wardle}, {Weintroub}, {Wex}, {Wharton}, {Wielgus}, {Wong}, {Wu}, {Yoon}, {Young}, {Young}, {Younsi}, {Yuan}, {Yuan}, {Zensus}, {Zhao}, {Zhao}, {Fermi Large Area Telescope Collaboration}, {Principe}, {Giroletti}, {D'Ammando}, {Orienti}, {H.~E.~S.~S. Collaboration}, {Abdalla}, {Adam}, {Aharonian}, {Benkhali}, {Ang{\"u}ner}, {Arcaro}, {Armand}, {Armstrong}, {Ashkar}, {Backes}, {Baghmanyan}, {Barbosa Martins}, {Barnacka},
  {Barnard}, {Becherini}, {Berge}, {Bernl{\"o}hr}, {Bi}, {B{\"o}ttcher}, {Boisson}, {Bolmont}, {de Lavergne}, {Breuhaus}, {Brun}, {Brun}, {Bryan}, {B{\"u}chele}, {Bulik}, {Bylund}, {Caroff}, {Carosi}, {Casanova}, {Chand}, {Chen}, {Cotter}, {Cury{\l}o}, {Damascene Mbarubucyeye}, {Davids}, {Davies}, {Deil}, {Devin}, {Dewilt}, {Dirson}, {Djannati-Ata{\"\i}}, {Dmytriiev}, {Donath}, {Doroshenko}, {Duffy}, {Dyks}, {Egberts}, {Eichhorn}, {Einecke}, {Emery}, {Ernenwein}, {Feijen}, {Fegan}, {Fiasson}, {de Clairfontaine}, {Fontaine}, {Funk}, {F{\"u}{\ss}ling}, {Gabici}, {Gallant}, {Giavitto}, {Giunti}, {Glawion}, {Glicenstein}, {Gottschall}, {Grondin}, {Hahn}, {Haupt}, {Hermann}, {Hinton}, {Hofmann}, {Hoischen}, {Holch}, {Holler}, {H{\"o}rbe}, {Horns}, {Huber}, {Jamrozy}, {Jankowsky}, {Jankowsky}, {Jardin-Blicq}, {Joshi}, {Jung-Richardt}, {Kasai}, {Kastendieck}, {Katarzy{\'n}ski}, {Katz}, {Khangulyan}, {Kh{\'e}lifi}, {Klepser}, {Klu{\'z}niak}, {Komin}, {Konno}, {Kosack}, {Kostunin}, {Kreter}, {Lamanna}, {Lemi{\`e}re},
  {Lemoine-Goumard}, {Lenain}, {Levy}, {Lohse}, {Lypova}, {Mackey}, {Majumdar}, {Malyshev}, {Malyshev}, {Marandon}, {Marchegiani}, {Marcowith}, {Mares}, {Mart{\'\i}-Devesa}, {Marx}, {Maurin}, {Meintjes}, {Meyer}, {Moderski}, {Mohamed}, {Mohrmann}, {Montanari}, {Moore}, {Morris}, {Moulin}, {Muller}, {Murach}, {Nakashima}, {Nayerhoda}, {de Naurois}, {Ndiyavala}, {Niederwanger}, {Niemiec}, {Oakes}, {O'Brien}, {Odaka}, {Ohm}, {Olivera-Nieto}, {de Ona Wilhelmi}, {Ostrowski}, {Panter}, {Panny}, {Parsons}, {Peron}, {Peyaud}, {Piel}, {Pita}, {Poireau}, {Noel}, {Prokhorov}, {Prokoph}, {P{\"u}hlhofer}, {Punch}, {Quirrenbach}, {Rauth}, {Reichherzer}, {Reimer}, {Reimer}, {Remy}, {Renaud}, {Rieger}, {Rinchiuso}, {Romoli}, {Rowell}, {Rudak}, {Ruiz-Velasco}, {Sahakian}, {Sailer}, {Sanchez}, {Santangelo}, {Sasaki}, {Scalici}, {Schutte}, {Schwanke}, {Schwemmer}, {Seglar-Arroyo}, {Senniappan}, {Seyffert}, {Shafi}, {Shiningayamwe}, {Simoni}, {Sinha}, {Sol}, {Specovius}, {Spencer}, {Spir-Jacob}, {Stawarz}, {Sun}, {Steenkamp},
  {Stegmann}, {Steinmassl}, {Steppa}, {Takahashi}, {Tavernier}, {Taylor}, {Terrier}, {Tiziani}, {Tluczykont}, {Tomankova}, {Trichard}, {Tsirou}, {Tuffs}, {Uchiyama}, {van der Walt}, {van Eldik}, {van Rensburg}, {van Soelen}, {Vasileiadis}, {Veh}, {Venter}, {Vincent}, {Vink}, {V{\"o}lk}, {Vuillaume}, {Wadiasingh}, {Wagner}, {Watson}, {Werner}, {White}, {Wierzcholska}, {Wong}, {Yusafzai}, {Zacharias}, {Zanin}, {Zargaryan}, {Zdziarski}, {Zech}, {Zhu}, {Zorn}, {Zouari}, {{\.Z}ywucka}, {MAGIC Collaboration}, {Acciari}, {Ansoldi}, {Antonelli}, {Engels}, {Artero}, {Asano}, {Baack}, {Babi{\'c}}, {Baquero}, {de Almeida}, {Barrio}, {Becerra Gonz{\'a}lez}, {Bednarek}, {Bellizzi}, {Bernardini}, {Bernardos}, {Berti}, {Besenrieder}, {Bhattacharyya}, {Bigongiari}, {Biland}, {Blanch}, {Bonnoli}, {Bo{\v{s}}njak}, {Busetto}, {Carosi}, {Ceribella}, {Cerruti}, {Chai}, {Chilingarian}, {Cikota}, {Colak}, {Colombo}, {Contreras}, {Cortina}, {Covino}, {D'Amico}, {D'Elia}, {da Vela}, {Dazzi}, {de Angelis}, {de Lotto}, {Delfino},
  {Delgado}, {Delgado Mendez}, {Depaoli}, {di Pierro}, {di Venere}, {Do Souto Espi{\~n}eira}, {Dominis Prester}, {Donini}, {Dorner}, {Doro}, {Elsaesser}, {Ramazani}, {Fattorini}, {Ferrara}, {Fonseca}, {Font}, {Fruck}, {Fukami}, {Garc{\'\i}a L{\'o}pez}, {Garczarczyk}, {Gasparyan}, {Gaug}, {Giglietto}, {Giordano}, {Gliwny}, {Godinovi{\'c}}, {Green}, {Green}, {Hadasch}, {Hahn}, {Heckmann}, {Herrera}, {Hoang}, {Hrupec}, {H{\"u}tten}, {Inada}, {Inoue}, {Ishio}, {Iwamura}, {Jim{\'e}nez}, {Jormanainen}, {Jouvin}, {Kajiwara}, {Karjalainen}, {Kerszberg}, {Kobayashi}, {Kubo}, {Kushida}, {Lamastra}, {Lelas}, {Leone}, {Lindfors}, {Lombardi}, {Longo}, {L{\'o}pez-Coto}, {L{\'o}pez-Moya}, {L{\'o}pez-Oramas}, {Loporchio}, {Machado de Oliveira Fraga}, {Maggio}, {Majumdar}, {Makariev}, {Mallamaci}, {Maneva}, {Manganaro}, {Mannheim}, {Maraschi}, {Mariotti}, {Mart{\'\i}nez}, {Mazin}, {Menchiari}, {Mender}, {Mi{\'c}anovi{\'c}}, {Miceli}, {Miener}, {Minev}, {Miranda}, {Mirzoyan}, {Molina}, {Moralejo}, {Morcuende}, {Moreno},
  {Moretti}, {Neustroev}, {Nigro}, {Nilsson}, {Nishijima}, {Noda}, {Nozaki}, {Ohtani}, {Oka}, {Otero-Santos}, {Paiano}, {Palatiello}, {Paneque}, {Paoletti}, {Paredes}, {Pavleti{\'c}}, {Pe{\~n}il}, {Perennes}, {Persic}, {Moroni}, {Prandini}, {Priyadarshi}, {Puljak}, {Rhode}, {Rib{\'o}}, {Rico}, {Righi}, {Rugliancich}, {Saha}, {Sahakyan}, {Saito}, {Sakurai}, {Satalecka}, {Saturni}, {Schleicher}, {Schmidt}, {Schweizer}, {Sitarek}, {{\v{S}}nidari{\'c}}, {Sobczynska}, {Spolon}, {Stamerra}, {Strom}, {Strzys}, {Suda}, {Suri{\'c}}, {Takahashi}, {Tavecchio}, {Temnikov}, {Terzi{\'c}}, {Teshima}, {Tosti}, {Truzzi}, {Tutone}, {Ubach}, {van Scherpenberg}, {Vanzo}, {Vazquez Acosta}, {Ventura}, {Verguilov}, {Vigorito}, {Vitale}, {Vovk}, {Will}, {Wunderlich}, {Zari{\'c}}, {VERITAS Collaboration}, {Adams}, {Benbow}, {Brill}, {Capasso}, {Christiansen}, {Chromey}, {Daniel}, {Errando}, {Farrell}, {Feng}, {Finley}, {Fortson}, {Furniss}, {Gent}, {Giuri}, {Hassan}, {Hervet}, {Holder}, {Hughes}, {Humensky}, {Jin}, {Kaaret},
  {Kertzman}, {Kieda}, {Kumar}, {Lang}, {Lundy}, {Maier}, {Moriarty}, {Mukherjee}, {Nieto}, {Nievas-Rosillo}, {O'Brien}, {Ong}, {Otte}, {Patel}, {Pfrang}, {Pohl}, {Prado}, {Pueschel}, {Quinn}, {Ragan}, {Reynolds}, {Ribeiro}, {Richards}, {Roache}, {Rulten}, {Ryan}, {Santander}, {Sembroski}, {Shang}, {Weinstein}, {Williams}, {Williamson}, {Eavn Collaboration}, {Hirota}, {Cui}, {Niinuma}, {Ro}, {Sakai}, {Sawada-Satoh}, {Wajima}, {Wang}, {Liu}, \& {Yonekura}}]{M87multi}
{EHT MWL Science Working Group}, {Algaba}, J.~C., {Anczarski}, J., {et~al.} 2021, \apjl, 911, L11

\bibitem[{{Feain} {et~al.}(2011){Feain}, {Cornwell}, {Ekers}, {Calabretta}, {Norris}, {Johnston-Hollitt}, {Ott}, {Lindley}, {Gaensler}, {Murphy}, {Middelberg}, {Jiraskova}, {O'Sullivan}, {McClure-Griffiths}, \& {Bland-Hawthorn}}]{Feain}
{Feain}, I.~J., {Cornwell}, T.~J., {Ekers}, R.~D., {et~al.} 2011, \apj, 740, 17

\bibitem[{{Fromm} {et~al.}(2016){Fromm}, {Perucho}, {Mimica}, \& {Ros}}]{Fromm2016}
{Fromm}, C.~M., {Perucho}, M., {Mimica}, P., \& {Ros}, E. 2016, Astronomy and Astrophysics, 588, A101

\bibitem[{{Gabuzda}(2018)}]{Gabuzda2018}
{Gabuzda}, D. 2018, Galaxies, 6, 9

\bibitem[{{Gabuzda} {et~al.}(2008){Gabuzda}, {Vitrishchak}, {Mahmud}, \& {O'Sullivan}}]{Gabuzda2008}
{Gabuzda}, D.~C., {Vitrishchak}, V.~M., {Mahmud}, M., \& {O'Sullivan}, S. 2008, Astronomical Society of the Pacific Conference Series, Vol. 386, Circular Polarization and Helical B Fields in AGN, 444

\bibitem[{{Ghisellini, Gabriele}(2013)}]{Ghisellini}
{Ghisellini, Gabriele}. 2013, EPJ Web of Conferences, 61, 05001

\bibitem[{{Giri} {et~al.}(2022){Giri}, {Dubey}, {Rubinur}, {Vaidya}, \& {Kharb}}]{Girib}
{Giri}, G., {Dubey}, R.~P., {Rubinur}, K., {Vaidya}, B., \& {Kharb}, P. 2022, \mnras, 514, 5625

\bibitem[{{Hada} {et~al.}(2011){Hada}, {Doi}, {Kino}, {Nagai}, {Hagiwara}, \& {Kawaguchi}}]{Hada11}
{Hada}, K., {Doi}, A., {Kino}, M., {et~al.} 2011, \nat, 477, 185

\bibitem[{{Hardcastle} {et~al.}(2003){Hardcastle}, {Worrall}, {Kraft}, {Forman}, {Jones}, \& {Murray}}]{Hardcastle}
{Hardcastle}, M.~J., {Worrall}, D.~M., {Kraft}, R.~P., {et~al.} 2003, \apj, 593, 169

\bibitem[{{Harris} {et~al.}(2010){Harris}, {Rejkuba}, \& {Harris}}]{Harris2010}
{Harris}, G. L.~H., {Rejkuba}, M., \& {Harris}, W.~E. 2010, Publications of the Astronomical Society of Australia, 27, 457

\bibitem[{{Heavens} \& {Meisenheimer}(1987)}]{Heavens}
{Heavens}, A.~F. \& {Meisenheimer}, K. 1987, \mnras, 225, 335

\bibitem[{{Hirotani} {et~al.}(1992){Hirotani}, {Takahashi}, {Nitta}, \& {Tomimatsu}}]{Hirotani}
{Hirotani}, K., {Takahashi}, M., {Nitta}, S.-Y., \& {Tomimatsu}, A. 1992, Astrophysical Journal, 386, 455

\bibitem[{{Homan} {et~al.}(2018){Homan}, {Hovatta}, {Kovalev}, {Lister}, {Pushkarev}, \& {Savolainen}}]{Homan}
{Homan}, D., {Hovatta}, T., {Kovalev}, Y., {et~al.} 2018, Galaxies, 6, 17

\bibitem[{{Janssen} {et~al.}(2021){Janssen}, {Falcke}, {Kadler}, {Ros}, {Wielgus}, {Akiyama}, {Balokovi{\'c}}, {Blackburn}, {Bouman}, {Chael}, {Chan}, {Chatterjee}, {Davelaar}, {Edwards}, {Fromm}, {G{\'o}mez}, {Goddi}, {Issaoun}, {Johnson}, {Kim}, {Koay}, {Krichbaum}, {Liu}, {Liuzzo}, {Markoff}, {Markowitz}, {Marrone}, {Mizuno}, {M{\"u}ller}, {Ni}, {Pesce}, {Ramakrishnan}, {Roelofs}, {Rygl}, {van Bemmel}, {Event Horizon Telescope Collaboration}, {Alberdi}, {Alef}, {Algaba}, {Anantua}, {Asada}, {Azulay}, {Baczko}, {Ball}, {Ball}, {Barrett}, {Benson}, {Bintley}, {Bintley}, {Blundell}, {Boland}, {Boland}, {Bower}, {Boyce}, {Bremer}, {Brinkerink}, {Brissenden}, {Britzen}, {Broderick}, {Broguiere}, {Bronzwaer}, {Byun}, {Carlstrom}, {Chatterjee}, {Chen}, {Chen}, {Chesler}, {Cho}, {Christian}, {Conway}, {Cordes}, {Crawford}, {Crew}, {Cruz-Osorio}, {Cui}, {Cui}, {De Laurentis}, {Deane}, {Dempsey}, {Desvignes}, {Dexter}, {Doeleman}, {Eatough}, {Farah}, {Farah}, {Fish}, {Fomalont}, {Ford}, {Fraga-Encinas}, {Friberg},
  {Friberg}, {Fuentes}, {Galison}, {Gammie}, {Garc{\'\i}a}, {Gelles}, {Gentaz}, {Georgiev}, {Georgiev}, {Gold}, {Gold}, {G{\'o}mez-Ruiz}, {Gu}, {Gurwell}, {Hada}, {Haggard}, {Hecht}, {Hesper}, {Himwich}, {Ho}, {Ho}, {Honma}, {Huang}, {Huang}, {Hughes}, {Ikeda}, {Inoue}, {Inoue}, {James}, {Jannuzi}, {Jeter}, {Jiang}, {Jimenez-Rosales}, {Jorstad}, {Jung}, {Karami}, {Karuppusamy}, {Kawashima}, {Keating}, {Kettenis}, {Kim}, {Kim}, {Kim}, {Kino}, {Kofuji}, {Koyama}, {Kramer}, {Kramer}, {Kuo}, {Lauer}, {Lee}, {Levis}, {Li}, {Li}, {Lindqvist}, {Lico}, {Lindahl}, {Liu}, {Lo}, {Lobanov}, {Loinard}, {Lonsdale}, {Lu}, {MacDonald}, {Mao}, {Marchili}, {Marscher}, {Mart{\'\i}-Vidal}, {Matsushita}, {Matthews}, {Medeiros}, {Menten}, {Mizuno}, {Moran}, {Moriyama}, {Moscibrodzka}, {Moscibrodzka}, {Musoke}, {Mej{\'\i}as}, {Nagai}, {Nagar}, {Nakamura}, {Narayan}, {Narayanan}, {Natarajan}, {Nathanail}, {Neilsen}, {Neri}, {Noutsos}, {Nowak}, {Okino}, {Olivares}, {Ortiz-Le{\'o}n}, {Oyama}, {{\"O}zel}, {Palumbo}, {Park}, {Patel},
  {Pen}, {Pi{\'e}tu}, {Plambeck}, {PopStefanija}, {Porth}, {P{\"o}tzl}, {Prather}, {Preciado-L{\'o}pez}, {Psaltis}, {Pu}, {Pu}, {Rao}, {Rawlings}, {Raymond}, {Rezzolla}, {Ricarte}, {Ripperda}, {Rogers}, {Rose}, {Roshanineshat}, {Rottmann}, {Roy}, {Ruszczyk}, {S{\'a}nchez}, {S{\'a}nchez-Arguelles}, {Sasada}, {Savolainen}, {Schloerb}, {Schuster}, {Shao}, {Shen}, {Small}, {Sohn}, {SooHoo}, {Sun}, {Tazaki}, {Tetarenko}, {Tiede}, {Tilanus}, {Titus}, {Torne}, {Trent}, {Traianou}, {Trippe}, {van Bemmel}, {van Langevelde}, {van Rossum}, {Wagner}, {Ward-Thompson}, {Wardle}, {Weintroub}, {Wex}, {Wharton}, {Wharton}, {Wong}, {Wu}, {Yoon}, {Young}, {Young}, {Younsi}, {Yuan}, {Yuan}, {Zensus}, {Zhao}, \& {Zhao}}]{JanssenCenA}
{Janssen}, M., {Falcke}, H., {Kadler}, M., {et~al.} 2021, Nature Astronomy, 5, 1017

\bibitem[{{Jones} \& {Odell}(1977)}]{JonesOdell}
{Jones}, T.~W. \& {Odell}, S.~L. 1977, Astronomy and Astrophysics, 61, 291

\bibitem[{{Kang} {et~al.}(1992){Kang}, {Jones}, \& {Ryu}}]{Kang}
{Kang}, H., {Jones}, T.~W., \& {Ryu}, D. 1992, \apj, 385, 193

\bibitem[{{Kim} {et~al.}(2023){Kim}, {Janssen}, {Krichbaum}, {Boccardi}, {MacDonald}, {Ros}, {Lobanov}, \& {Zensus}}]{Kim2023}
{Kim}, D.-W., {Janssen}, M., {Krichbaum}, T.~P., {et~al.} 2023, \aap, 680, L3

\bibitem[{{Kim} {et~al.}(2019){Kim}, {Krichbaum}, {Marscher}, {Jorstad}, {Agudo}, {Thum}, {Hodgson}, {MacDonald}, {Ros}, {Lu}, {Bremer}, {de Vicente}, {Lindqvist}, {Trippe}, \& {Zensus}}]{Kim2019}
{Kim}, J.-Y., {Krichbaum}, T.~P., {Marscher}, A.~P., {et~al.} 2019, Astronomy and Astrophysics, 622, A196

\bibitem[{{Kramer} \& {MacDonald}(2021)}]{Kramer&MacDonald}
{Kramer}, J.~A. \& {MacDonald}, N.~R. 2021, \aap, 656, A143

\bibitem[{Laing {et~al.}(1999)Laing, Parma, Ruiter, \& Fanti}]{Laing99}
Laing, R.~A., Parma, P., Ruiter, H. R.~d., \& Fanti, R. 1999, Monthly Notices of the Royal Astronomical Society, 306, 513

\bibitem[{{Lister} \& {Homan}(2005)}]{Lister2005}
{Lister}, M.~L. \& {Homan}, D.~C. 2005, \aj, 130, 1389

\bibitem[{{Lobanov}(1998)}]{Lobanov98}
{Lobanov}, A.~P. 1998, \aaps, 132, 261

\bibitem[{{MacDonald} \& {Marscher}(2018)}]{MacDonald2018}
{MacDonald}, N.~R. \& {Marscher}, A.~P. 2018, Astrophysical Journal, 862, 58

\bibitem[{{MacDonald} \& {Nishikawa}(2021)}]{Macdonald2021}
{MacDonald}, N.~R. \& {Nishikawa}, K.~I. 2021, \aap, 653, A10

\bibitem[{{Mandal} {et~al.}(2023){Mandal}, {Mukherjee}, \& {Mignone}}]{Mandal2023}
{Mandal}, A., {Mukherjee}, D., \& {Mignone}, A. 2023, \apjs, 268, 40

\bibitem[{{Mattia} {et~al.}(2023){Mattia}, {Del Zanna}, {Bugli}, {Pavan}, {Ciolfi}, {Bodo}, \& {Mignone}}]{Mattia}
{Mattia}, G., {Del Zanna}, L., {Bugli}, M., {et~al.} 2023, \aap, 679, A49

\bibitem[{{Meisenheimer} \& {Roeser}(1986)}]{Meisenheimer}
{Meisenheimer}, K. \& {Roeser}, H.~J. 1986, Nature, 319, 459

\bibitem[{{Melon Fuksman} \& {Mignone}(2019)}]{Melon2019}
{Melon Fuksman}, J.~D. \& {Mignone}, A. 2019, \apjs, 242, 20

\bibitem[{{Mignone} {et~al.}(2007){Mignone}, {Bodo}, {Massaglia}, {Matsakos}, {Tesileanu}, {Zanni}, \& {Ferrari}}]{Mignone2007}
{Mignone}, A., {Bodo}, G., {Massaglia}, S., {et~al.} 2007, \apjs, 170, 228

\bibitem[{{Mignone} {et~al.}(2018){Mignone}, {Bodo}, {Vaidya}, \& {Mattia}}]{Mignone2018}
{Mignone}, A., {Bodo}, G., {Vaidya}, B., \& {Mattia}, G. 2018, \apj, 859, 13

\bibitem[{{Mignone} {et~al.}(2019){Mignone}, {Flock}, \& {Vaidya}}]{Mignone2019}
{Mignone}, A., {Flock}, M., \& {Vaidya}, B. 2019, \apjs, 244, 38

\bibitem[{{Mignone} {et~al.}(2023){Mignone}, {Haudemand}, \& {Puzzoni}}]{Mignone2023}
{Mignone}, A., {Haudemand}, H., \& {Puzzoni}, E. 2023, Computer Physics Communications, 285, 108625

\bibitem[{{Mimica} {et~al.}(2012){Mimica}, {Giannios, D.}, {Metzger, B.}, \& {Aloy, M.A.}}]{Mimica2012}
{Mimica}, {Giannios, D.}, {Metzger, B.}, \& {Aloy, M.A.} 2012, EPJ Web of Conferences, 39, 04003

\bibitem[{{Mimica} {et~al.}(2009){Mimica}, {Aloy}, {Agudo}, {Mart{\'\i}}, {G{\'o}mez}, \& {Miralles}}]{Mimica2009}
{Mimica}, P., {Aloy}, M.-A., {Agudo}, I., {et~al.} 2009, Astrophysical Journal, 696, 1142

\bibitem[{{Mizuno} {et~al.}(2015){Mizuno}, {G{\'o}mez}, {Nishikawa}, {Meli}, {Hardee}, \& {Rezzolla}}]{Mizuno2015}
{Mizuno}, Y., {G{\'o}mez}, J.~L., {Nishikawa}, K.-I., {et~al.} 2015, Astrophysical Journal, 809, 38

\bibitem[{{Mukherjee} {et~al.}(2021){Mukherjee}, {Bodo}, {Rossi}, {Mignone}, \& {Vaidya}}]{Mukherjee21}
{Mukherjee}, D., {Bodo}, G., {Rossi}, P., {Mignone}, A., \& {Vaidya}, B. 2021, \mnras, 505, 2267

\bibitem[{{M{\"u}ller} {et~al.}(2014){M{\"u}ller}, {Kadler}, {Ojha}, {Perucho}, {Gro{\ss}berger}, {Ros}, {Wilms}, {Blanchard}, {B{\"o}ck}, {Carpenter}, {Dutka}, {Edwards}, {Hase}, {Horiuchi}, {Kreikenbohm}, {Lovell}, {Markowitz}, {Phillips}, {Pl{\"o}tz}, {Pursimo}, {Quick}, {Rothschild}, {Schulz}, {Steinbring}, {Stevens}, {Tr{\"u}stedt}, \& {Tzioumis}}]{muller2014}
{M{\"u}ller}, C., {Kadler}, M., {Ojha}, R., {et~al.} 2014, Astronomy and Astrophysics, 569, A115

\bibitem[{{M{\"u}ller} {et~al.}(2011){M{\"u}ller}, {Kadler}, {Ojha}, {Wilms}, {B{\"o}ck}, {Edwards}, {Fromm}, {Hase}, {Horiuchi}, {Katz}, {Lovell}, {Pl{\"o}tz}, {Pursimo}, {Richers}, {Ros}, {Rothschild}, {Taylor}, {Tingay}, \& {Zensus}}]{Muller2011}
{M{\"u}ller}, C., {Kadler}, M., {Ojha}, R., {et~al.} 2011, \aap, 530, L11

\bibitem[{{Neumayer}(2010)}]{Neumayer}
{Neumayer}, N. 2010, Publications of the Astronomical Society of Australia, 27, 449

\bibitem[{{Nurisso} {et~al.}(2023){Nurisso}, {Celotti}, {Mignone}, \& {Bodo}}]{Nurisso2023}
{Nurisso}, M., {Celotti}, A., {Mignone}, A., \& {Bodo}, G. 2023, \mnras, 522, 5517

\bibitem[{{Ojha} {et~al.}(2010){Ojha}, {Kadler}, {B{\"o}ck}, {Booth}, {Dutka}, {Edwards}, {Fey}, {Fuhrmann}, {Gaume}, {Hase}, {Horiuchi}, {Jauncey}, {Johnston}, {Katz}, {Lister}, {Lovell}, {M{\"u}ller}, {Pl{\"o}tz}, {Quick}, {Ros}, {Taylor}, {Thompson}, {Tingay}, {Tosti}, {Tzioumis}, {Wilms}, \& {Zensus}}]{Ojha2010}
{Ojha}, R., {Kadler}, M., {B{\"o}ck}, M., {et~al.} 2010, \aap, 519, A45

\bibitem[{{Paraschos} {et~al.}(2024){Paraschos}, {Debbrecht}, {Kramer}, {Traianou}, {Liodakis}, {Krichbaum}, {Kim}, {Janssen}, {Nair}, {Savolainen}, {Ros}, {Bach}, {Hodgson}, {Lisakov}, {MacDonald}, \& {Zensus}}]{Paraschos24}
{Paraschos}, G.~F., {Debbrecht}, L.~C., {Kramer}, J.~A., {et~al.} 2024, \aap, 686, L5

\bibitem[{{Paraschos} {et~al.}(2021){Paraschos}, {Kim}, {Krichbaum}, \& {Zensus}}]{Paraschos}
{Paraschos}, G.~F., {Kim}, J.~Y., {Krichbaum}, T.~P., \& {Zensus}, J.~A. 2021, \aap, 650, L18

\bibitem[{{Paraschos} {et~al.}(2023){Paraschos}, {Mpisketzis}, {Kim}, {Witzel}, {Krichbaum}, {Zensus}, {Gurwell}, {L{\"a}hteenm{\"a}ki}, {Tornikoski}, {Kiehlmann}, \& {Readhead}}]{Paraschos23}
{Paraschos}, G.~F., {Mpisketzis}, V., {Kim}, J.~Y., {et~al.} 2023, \aap, 669, A32

\bibitem[{{Park} \& {Algaba}(2022)}]{Park2020}
{Park}, J. \& {Algaba}, J.~C. 2022, Galaxies, 10, 102

\bibitem[{{Perucho} {et~al.}(2023){Perucho}, {L{\'o}pez-Miralles}, {Gizani}, {Mart{\'\i}}, \& {Boccardi}}]{Perucho2023}
{Perucho}, M., {L{\'o}pez-Miralles}, J., {Gizani}, N. A.~B., {Mart{\'\i}}, J.~M., \& {Boccardi}, B. 2023, \mnras, 523, 3583

\bibitem[{{Pontin} \& {Priest}(2022)}]{magrecon}
{Pontin}, D.~I. \& {Priest}, E.~R. 2022, Living Reviews in Solar Physics, 19, 1

\bibitem[{{Porth} {et~al.}(2011){Porth}, {Fendt}, {Meliani}, \& {Vaidya}}]{Porth2011}
{Porth}, O., {Fendt}, C., {Meliani}, Z., \& {Vaidya}, B. 2011, Astrophysical Journal, 737, 42

\bibitem[{{Pushkarev} {et~al.}(2017){Pushkarev}, {Kovalev}, {Lister}, {Savolainen}, {Aller}, {Aller}, \& {Hodge}}]{Pushkarev2017}
{Pushkarev}, A., {Kovalev}, Y., {Lister}, M., {et~al.} 2017, Galaxies, 5, 93

\bibitem[{{Pushkarev} {et~al.}(2005){Pushkarev}, {Gabuzda}, {Vetukhnovskaya}, \& {Yakimov}}]{Pushkarev2005}
{Pushkarev}, A.~B., {Gabuzda}, D.~C., {Vetukhnovskaya}, Y.~N., \& {Yakimov}, V.~E. 2005, \mnras, 356, 859

\bibitem[{{Pushkarev} {et~al.}(2012){Pushkarev}, {Hovatta}, {Kovalev}, {Lister}, {Lobanov}, {Savolainen}, \& {Zensus}}]{Pushkarev2012}
{Pushkarev}, A.~B., {Hovatta}, T., {Kovalev}, Y.~Y., {et~al.} 2012, \aap, 545, A113

\bibitem[{{Rees}(1966)}]{Rees}
{Rees}, A.~I. 1966, Journal of Geology, 74, 856

\bibitem[{Rothschild {et~al.}(2011)Rothschild, Markowitz, Rivers, Suchy, Pottschmidt, Kadler, Müller, \& Wilms}]{CenAredshift}
Rothschild, R.~E., Markowitz, A., Rivers, E., {et~al.} 2011, The Astrophysical Journal, 733, 23

\bibitem[{{Rybicki} \& {Lightman}(1979)}]{Rybicki79}
{Rybicki}, G.~B. \& {Lightman}, A.~P. 1979, {Radiative processes in astrophysics}, 1st edn. (John Wiley \& Sons)

\bibitem[{{Sokolovsky} {et~al.}(2011){Sokolovsky}, {Kovalev}, {Pushkarev}, \& {Lobanov}}]{89}
{Sokolovsky}, K.~V., {Kovalev}, Y.~Y., {Pushkarev}, A.~B., \& {Lobanov}, A.~P. 2011, A\&A, 532, A38

\bibitem[{{Sow Mondal} {et~al.}(2023){Sow Mondal}, {Sarkar}, {Vaidya}, \& {Mignone}}]{Mondal2023}
{Sow Mondal}, S., {Sarkar}, A., {Vaidya}, B., \& {Mignone}, A. 2023, in AAS/Solar Physics Division Meeting, Vol.~55, 54th Meeting of the Solar Physics Division, 106.30

\bibitem[{{Tingay} \& {Murphy}(2001)}]{Tingay}
{Tingay}, S.~J. \& {Murphy}, D.~W. 2001, \apj, 546, 210

\bibitem[{{Todorov} {et~al.}(2023){Todorov}, {Kravchenko}, {Pashchenko}, \& {Pushkarev}}]{Todorov}
{Todorov}, R.~V., {Kravchenko}, E.~V., {Pashchenko}, I.~N., \& {Pushkarev}, A.~B. 2023, Astronomy Reports, 67, 1275

\bibitem[{{Tsunetoe} {et~al.}(2021){Tsunetoe}, {Mineshige}, {Ohsuga}, {Kawashima}, \& {Akiyama}}]{Tsunetoe}
{Tsunetoe}, Y., {Mineshige}, S., {Ohsuga}, K., {Kawashima}, T., \& {Akiyama}, K. 2021, \pasj, 73, 912

\bibitem[{{Urry} \& {Padovani}(1995)}]{Urry1995}
{Urry}, C.~M. \& {Padovani}, P. 1995, Publications of the Astronomical Society of the Pacific, 107, 803

\bibitem[{{Vaidya} {et~al.}(2016){Vaidya}, {Mignone}, {Bodo}, \& {Massaglia}}]{Vaidya2016}
{Vaidya}, B., {Mignone}, A., {Bodo}, G., \& {Massaglia}, S. 2016, in Journal of Physics Conference Series, Vol. 719, Journal of Physics Conference Series, 012023

\bibitem[{{Vaidya} {et~al.}(2018){Vaidya}, {Mignone}, {Bodo}, {Rossi}, \& {Massaglia}}]{Vaidya2018}
{Vaidya}, B., {Mignone}, A., {Bodo}, G., {Rossi}, P., \& {Massaglia}, S. 2018, Astrophysical Journal, 865, 144

\bibitem[{{van Marle} {et~al.}(2018){van Marle}, {Casse}, \& {Marcowith}}]{Marle}
{van Marle}, A.~J., {Casse}, F., \& {Marcowith}, A. 2018, \mnras, 473, 3394

\bibitem[{{Wardle} {et~al.}(1998){Wardle}, {Homan}, {Ojha}, \& {Roberts}}]{Wardle}
{Wardle}, J.~F.~C., {Homan}, D.~C., {Ojha}, R., \& {Roberts}, D.~H. 1998, Nature, 395, 457

\bibitem[{{Webb}(1989)}]{Webb}
{Webb}, G.~M. 1989, \apj, 340, 1112

\bibitem[{{Worrall} {et~al.}(2008){Worrall}, {Birkinshaw}, {Kraft}, {Sivakoff}, {Jord{\'a}n}, {Hardcastle}, {Brassington}, {Croston}, {Evans}, {Forman}, {Harris}, {Jones}, {Juett}, {Murray}, {Nulsen}, {Raychaudhury}, {Sarazin}, \& {Woodley}}]{Worrall2008}
{Worrall}, D.~M., {Birkinshaw}, M., {Kraft}, R.~P., {et~al.} 2008, \apjl, 673, L135

\end{thebibliography}

\begin{appendix}
\section{Structure}\label{app:structure}
\begin{figure}[h!]
    \centering
    \includegraphics[trim={    0.8cm     0.0cm      1.75cm       0.0cm      }, clip,width=0.48\textwidth]{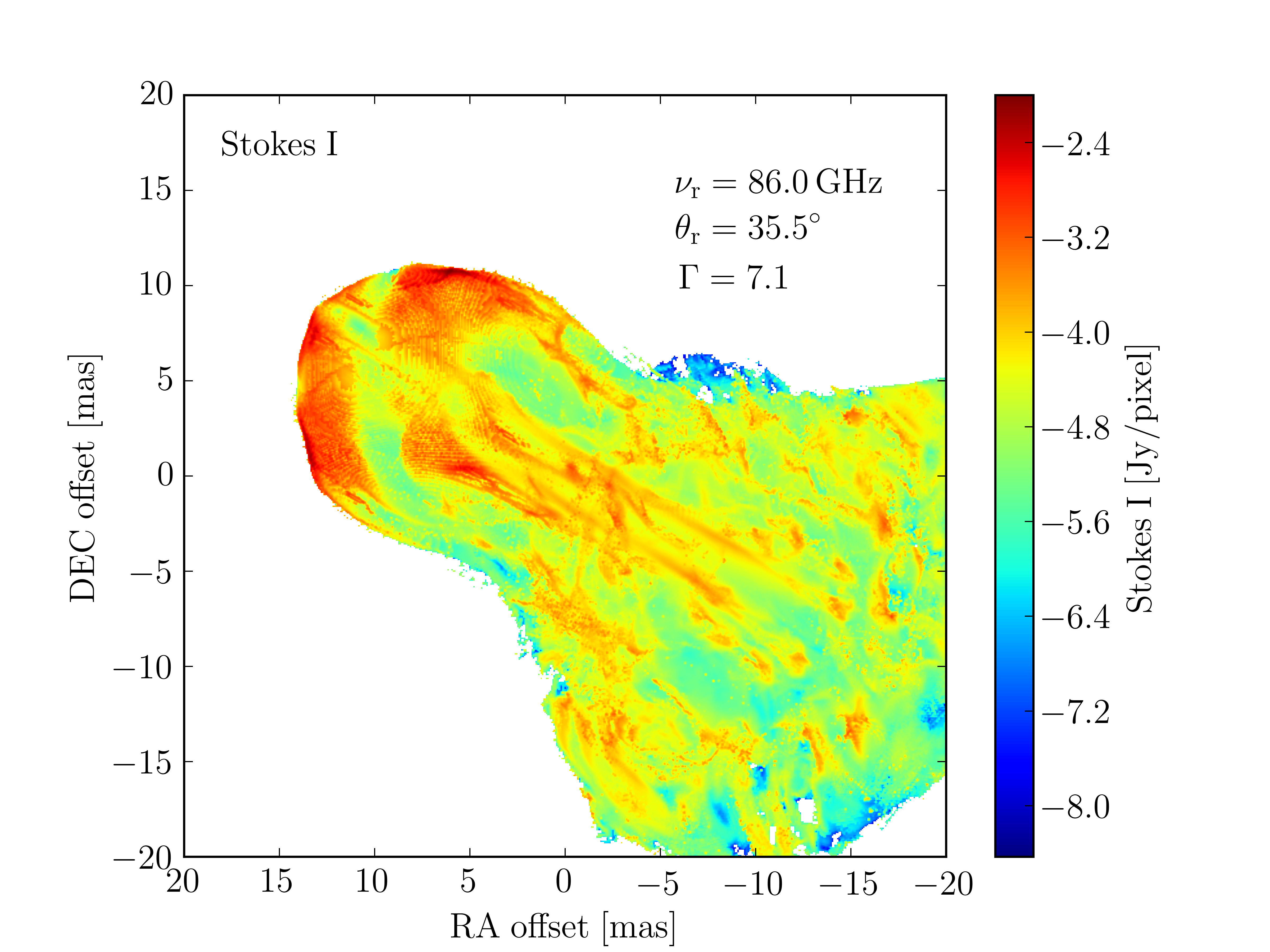}
    \caption{Resolved $86\,$GHz total intensity maps of the 3D RMHD jet simulation presented in \cite{Kramer&MacDonald} in logarithmic color-coding. The jet emission is  viewed at an observational viewing angle of 35.5$^\circ$. The logarithmic plot reveals an edge-brightening jet hidden behind the bow shock.}
    \label{fig:log}
\end{figure}

An underlying structure comes to light when employing a logarithmic scale in our analysis: the edge-brightened jet structure is distinct and can be indeed observed through the dense cocoon, as illustrated in Fig.~\ref{fig:log}. This figure shows a comprehensive view of a total intensity map derived from a preceding 3D RMHD jet simulation~\citep{Kramer&MacDonald}. It is noteworthy that this outcome is achieved without the use of any specialized jet tracer or intricate particle physics mechanisms. One prominent feature highlighted in the figure is the bow shock, which, under these circumstances, completely overshadows the jet structure. However, by considering radiative losses, a physical effect takes place. The bow shock, which comprises particles flowing in reverse, undergoes a cooling process, leading to a reduction in its intensity and overall prominence. This insight stresses the complex interplay between various physical factors that ultimately shape the observable characteristics of the jet and its surrounding environment. Figure~\ref{fig:log} proves that we are indeed using the very same numerical setup, and that we can obtain the result of an unobscured edge-brightened jet when we choose the hybrid fluid-particle approach. 
\section{Particle shock acceleration}
\begin{figure}[h!]
    \centering
    \includegraphics[trim={    0.8cm     0.0cm      1.75cm       0.0cm      }, clip,width=0.48\textwidth]{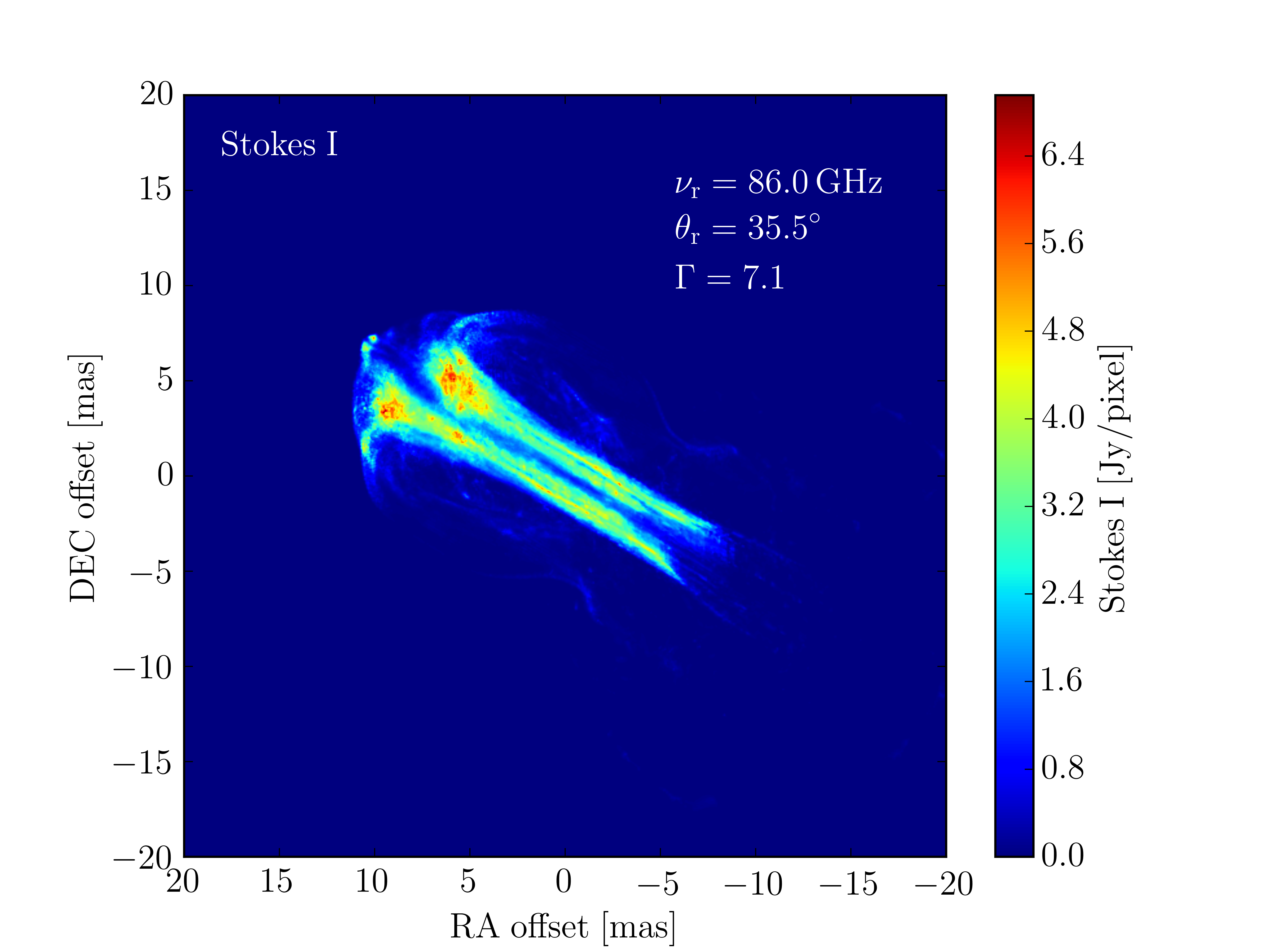}
    \caption{Resolved $86\,$GHz total intensity maps of the 3D hybrid fluid-particle jet simulation with the inclusion of particle shock acceleration. The jet emission is  viewed at an observational viewing angle of 35.5$^\circ$. The inclusion of acceleration mechanisms reveals the highlighted bow shock in addition to the edge-brightened jet.}
    \label{app:shock}
\end{figure}

Although our paper focuses on the importance of radiative losses through the inclusion of particle physics, the impact of particle shock acceleration is also significant. This form of acceleration is crucial in the presence of phenomena like a recollimation shock, a plasmoid moving through the jet, magnetic reconnection events, or instabilities~\citep{Borse, Mukherjee21}. It is noteworthy that our initialization of the spectral index value, namely $\alpha=2$, mimics a flat spectrum and does not enhance particle acceleration (flattening of the spectrum) too strongly. We chose this value based on radio observations~\citep[see, e.g.,][]{Paraschos}. Furthermore, our simulations, which utilize a purely toroidal magnetic field, do not exhibit any features that would significantly influence particle acceleration, as discussed in~\cite{Kramer&MacDonald}. The synthetic polarized emission strengthens toward the jet edges, suggesting that particles cluster around the toroidal magnetic field lines. Figure~\ref{app:shock} displays structures similar to those in Fig.~\ref{fig:comp}, notably an edge-brightened jet structure. Including particle acceleration mechanisms such as diffusive shock acceleration would further highlight the jet emission, especially emphasizing the strong bow shock moving through the ambient medium~\citep{Girib}. We also confirm higher values of total intensity due to the presence of the strong bow shock, visible in Fig.~\ref{fig:comp} as well. For a more detailed discussion on shock acceleration, we refer the reader to studies like those by, for example, \cite{Dubey2023}, \cite{Mattia}, and \cite{Girib}.

\section{Frequency}\label{app:freq}
\begin{figure}
    \begin{tikzpicture}
    \node[anchor=north west,inner sep=0] (Fig1) at (0,0) 
    {\includegraphics[trim={    0.5cm     0.0cm      1.9cm       1.0cm      }, clip,width=0.48\textwidth]{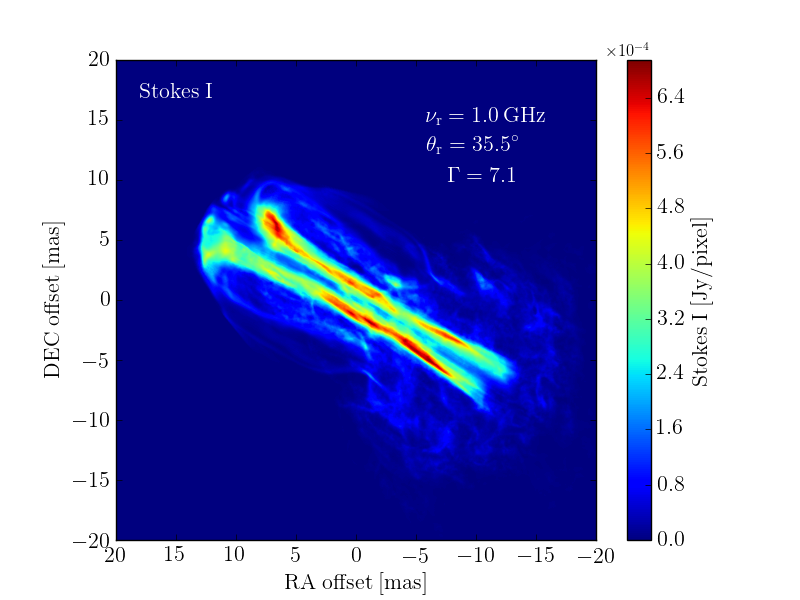}};
    \node[anchor=north west, inner sep=0] (Fig2) at (Fig1.south west) 
    {\includegraphics[trim={    0.5cm     0.0cm      2.1cm       1.0cm      }, clip,width=0.48\textwidth]{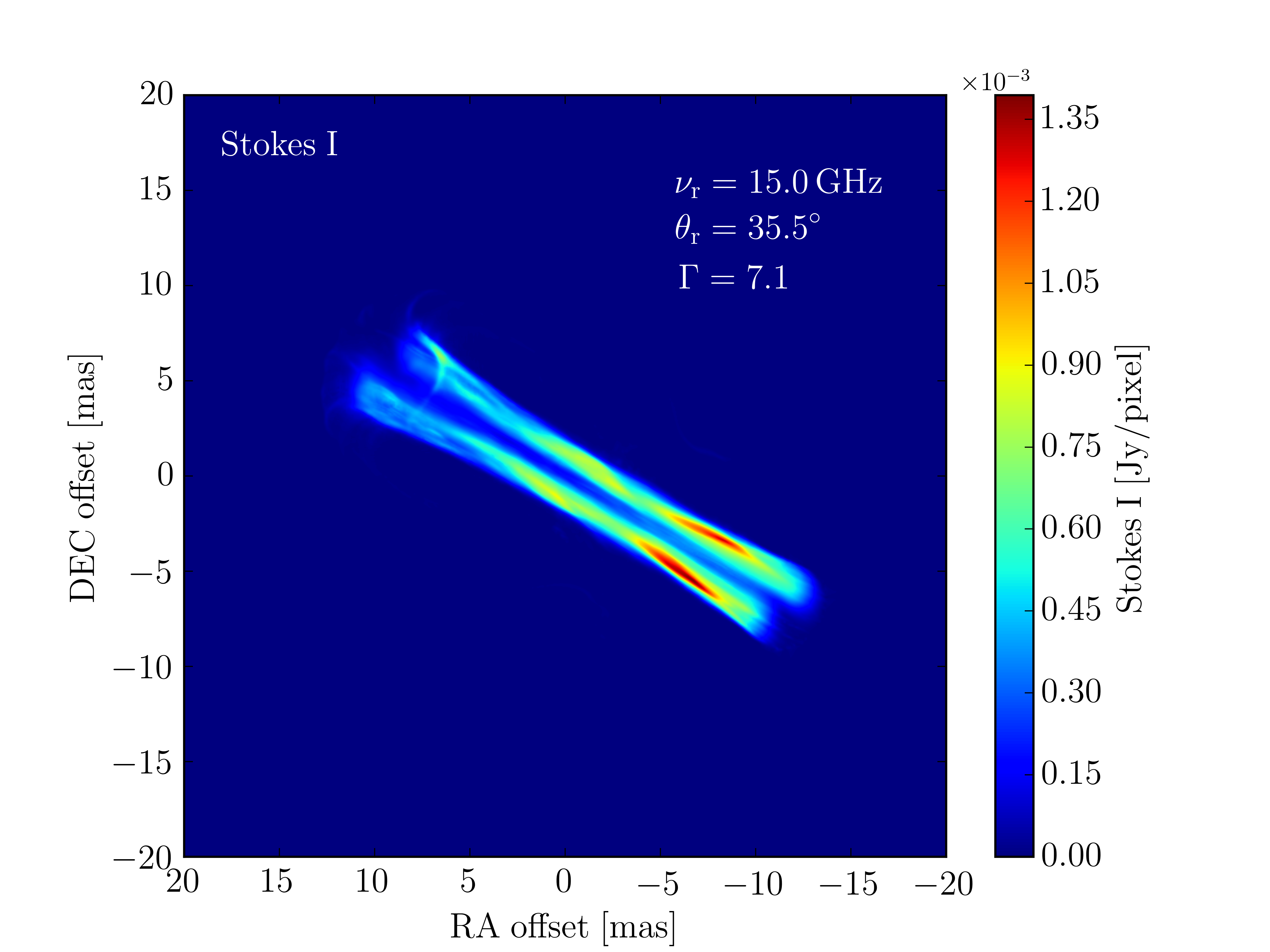}};
    \node[anchor=north west, inner sep=0] (Fig3) at (Fig2.south west) 
    {\includegraphics[trim={    0.5cm     0.0cm      2.1cm       1.0cm      }, clip,width=0.48\textwidth]{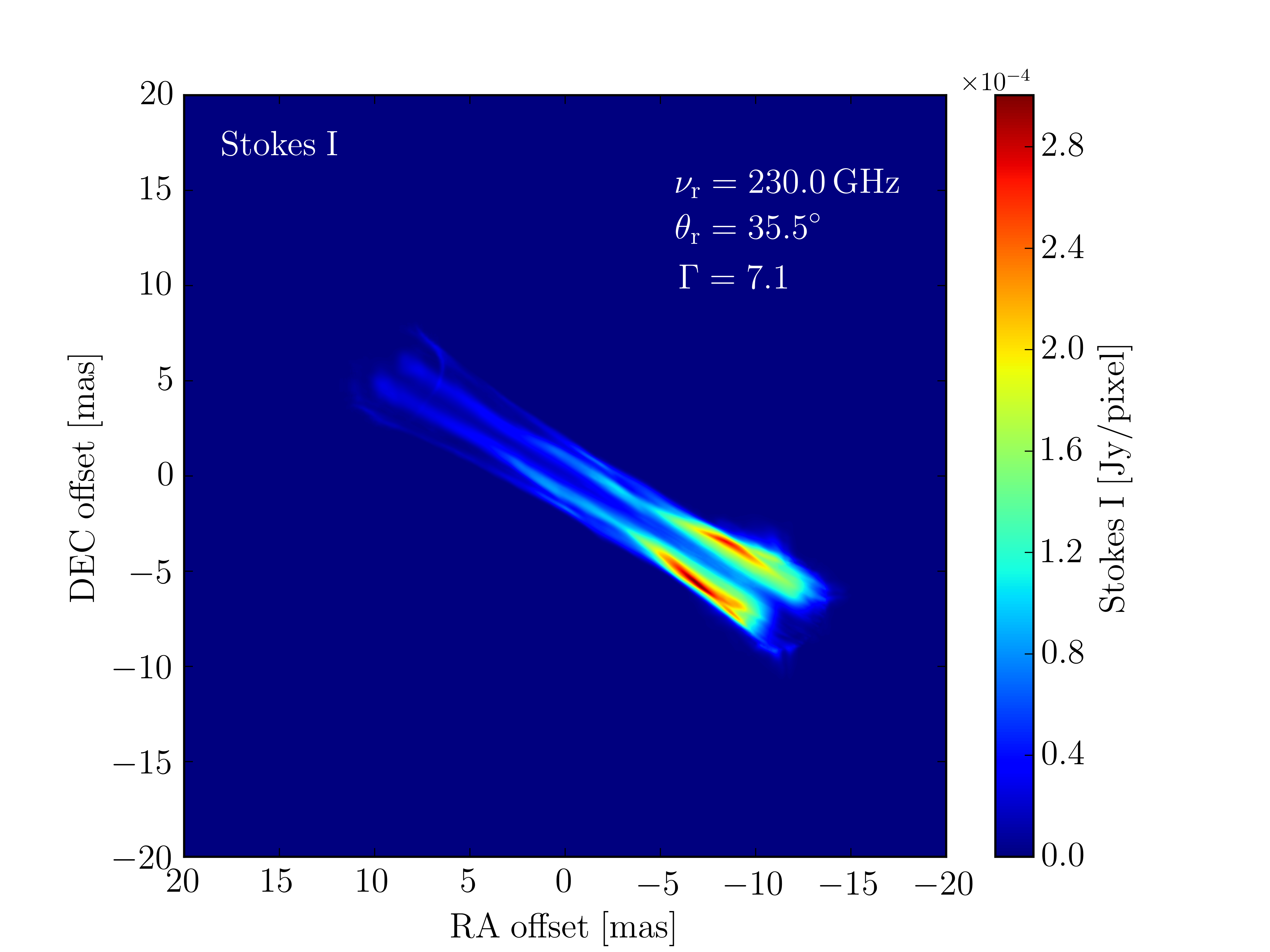}};
    \end{tikzpicture}
    \caption{Resolved total intensity maps of the final 3D hybrid fluid-particle jet simulation revealing an edge-brightened jet emission in 1\,GHz (\textbf{top}), 15\,GHz (\textbf{middle}), and 230\,GHz (\textbf{bottom}). All three maps are viewed at an observational viewing angle of 35.5$^\circ$. The figure highlights the influence opacity effects have on resultant synthetic synchrotron emission.}
    \label{fig:app}
\end{figure}

Our observations of the polarized total intensity maps of synthetic synchrotron emission shown in Fig.~\ref{fig:app} reveal several key characteristics of the jet's emission. Specifically, the emission pattern displays edge-brightening, an effect that can be attributed to the presence of a toroidal magnetic field configuration~\citep{Laing99}.
Notably, this distinctive feature highlights the physical insights into jet collimation that can be gained from RMHD simulations~\citep{Gabuzda2018,Pushkarev2005}.

Our analysis also reveals an opacity effect within the compact region. 
This effect is commonly referred to as "core shift" and is a phenomenon manifested as the apparent shift of the unresolved compact region toward the central engine, with increasing observing frequency~\citep{Lobanov98, Hirotani, Hada11, Pushkarev2005, Pushkarev2017, Paraschos, Park2020, Paraschos23}.
Such higher frequency observations can peer deeper into synchrotron self-absorbed sources, as they then appear less opaque.
Below we distinguish between two cases (case I and II) for the nature of the ``radio core.''

In case I we defined the unresolved region as the radio core at the peak brightness point of the emission. 
We detect a discernible shift of the radio core position toward the central engine when moving to higher frequencies (from 1\,GHz to 15\,GHz and 86\,GHz to 230\,GHz). 

Alternatively, in case II, the radio core is considered to be the boundary between regions of upstream synchrotron self-absorption and downstream optically thin segments; then a compelling argument emerges. 
In this context, identifying the radio core with the location of the recollimation shock becomes a plausible proposition, as can be seen in the most left panel in Fig.~\ref{fig:pipeline}.
If that scenario is correct, studying radio cores with very long baseline interferometry  can reveal new insights into jet launching.

Finally, a noteworthy trend comes to light in our simulations: the jet's observed brightness progressively diminishes, as we ascend toward higher frequencies.
This suggests that our analysis of the jet is in the optically thin regime, beyond the turnover frequency.
As the observing capabilities of state-of-the-art arrays like the EHT expand into ever higher frequencies~\citep[see, e.g.,][]{Crew}, insights into the jet physics in the optically thin regime will be crucial.
Simulations like the ones presented here are primed to advance our understanding of the nature of jets.

\end{appendix}



\end{document}